%% file: main.tex
\PassOptionsToPackage{dvipsnames}{xcolor}
\documentclass[11pt]{article}
\usepackage{jheppub}
\usepackage{amsthm}
\usepackage{tikz}
\usepackage{circuitikz}
\usetikzlibrary{decorations.pathreplacing,decorations.markings}
\usepackage{mathrsfs}
\usepackage{mathtools}
\usepackage{dsfont}
\usepackage{blkarray}
\usepackage{todonotes}
\usepackage[T1]{fontenc}
\usepackage{setspace}
\usepackage{subcaption}
\usepackage{braket}
\usepackage{multirow}
\usepackage{cancel}
\usepackage{adjustbox}
\usepackage{cleveref}
\usepackage{bm}
\usepackage{physics}
\usepackage{listings}
\usepackage{bbding}
\usepackage{parskip} 
\usepackage{fontawesome5}

\usepackage{amsmath,amssymb,shuffle,paralist,accents}
\usepackage{caption}
\usepackage{xcolor}
\usetikzlibrary{backgrounds, fit}

\usetikzlibrary{shapes.misc}
\usetikzlibrary{cd}
\usetikzlibrary{snakes}
\usetikzlibrary{calc}
\usetikzlibrary{fadings}
\usetikzlibrary{decorations.markings}
\usetikzlibrary{positioning}
\usetikzlibrary{shapes.multipart}

\usepackage{tabularx}
\usepackage{makecell}
\usepackage{mdframed}

\definecolor{truecolor}{RGB}{0,90,180}
\definecolor{falsecolor}{RGB}{210,100,0}

\newcommand{\True}{\texttt{\textcolor{truecolor}{True}}}
\newcommand{\False}{\texttt{\textcolor{falsecolor}{False}}}

\usepackage[most]{tcolorbox}

\newcommand{\bubble}{\text{\tikz[
    baseline=-0.6ex,
    line width=0.45pt,
    x=1em,
    y=1em
]{
    \draw (-0.34,0) .. controls (-0.13,0.18) and (0.13,0.18) .. (0.34,0);
    \draw (-0.34,0) .. controls (-0.13,-0.18) and (0.13,-0.18) .. (0.34,0);
}}}

\newcommand{\hooktriangle}{%
    \mathord{\vcenter{\hbox{%
        \tikz[
            line cap=round,
            line join=round,
            x=0.75em,
            y=0.75em
        ]{
            \coordinate (L) at (-0.30,0);
            \coordinate (U) at (0.22,0.30);
            \coordinate (D) at (0.22,-0.30);

            \draw[line width=0.45pt] (L) -- (U);

            \draw[line width=1.05pt] (L) -- (D);
            \draw[line width=1.05pt] (D) -- (U);
        }%
    }}}%
}

\newcommand{\masslessbox}{%
    \mathord{\vcenter{\hbox{%
        \tikz[
            baseline=-0.6ex,
            line cap=round,
            line join=round,
            x=0.55em,
            y=0.55em
        ]{
            \coordinate (TL) at (-0.28,  0.28);
            \coordinate (TR) at ( 0.28,  0.28);
            \coordinate (BR) at ( 0.28, -0.28);
            \coordinate (BL) at (-0.28, -0.28);

            \draw[line width=0.45pt] (TL) -- (TR);
            \draw[line width=0.45pt] (TR) -- (BR);
            \draw[line width=0.45pt] (BR) -- (BL);
            \draw[line width=0.45pt] (BL) -- (TL);

            \draw[line width=0.45pt] (TL) -- (-0.53,  0.53);
            \draw[line width=0.45pt] (TR) -- ( 0.53,  0.53);
            \draw[line width=0.45pt] (BR) -- ( 0.53, -0.53);
            \draw[line width=0.45pt] (BL) -- (-0.53, -0.53);
        }%
    }}}%
}

\newcommand{\twomasseasybox}{%
    \mathord{\vcenter{\hbox{%
        \tikz[
            baseline=-0.6ex,
            line cap=round,
            line join=round,
            x=0.55em,
            y=0.55em
        ]{
            \coordinate (TL) at (-0.28,  0.28);
            \coordinate (TR) at ( 0.28,  0.28);
            \coordinate (BR) at ( 0.28, -0.28);
            \coordinate (BL) at (-0.28, -0.28);

            \draw[line width=0.45pt] (TL) -- (TR);
            \draw[line width=0.45pt] (TR) -- (BR);
            \draw[line width=0.45pt] (BR) -- (BL);
            \draw[line width=0.45pt] (BL) -- (TL);

            \draw[line width=0.45pt] (TL) -- (-0.58,  0.43); 
            \draw[line width=0.45pt] (TL) -- (-0.43,  0.62); 
            \draw[line width=0.45pt] (TR) -- ( 0.52,  0.52); 
            \draw[line width=0.45pt] (BR) -- ( 0.58, -0.43); 
            \draw[line width=0.45pt] (BR) -- ( 0.43, -0.62); 
            \draw[line width=0.45pt] (BL) -- (-0.53, -0.53); 
        }%
    }}}%
}

\newcommand{\FeynmanBubble}{%
    \mathord{\vcenter{\hbox{%
        \tikz[
            baseline=-0.6ex,
            line cap=round,
            line join=round,
            x=0.72em,
            y=0.73em
        ]{
            \coordinate (L) at (-0.24, 0);
            \coordinate (R) at ( 0.24, 0);

            \draw[line width=0.45pt] (L) .. controls (0, 0.18) .. (R);
            \draw[line width=0.45pt] (L) .. controls (0,-0.18) .. (R);

            \draw[line width=0.45pt] (L) -- (-0.50,  0.10);
            \draw[line width=0.45pt] (L) -- (-0.50, -0.10);
            \draw[line width=0.45pt] (R) -- ( 0.50,  0.10);
            \draw[line width=0.45pt] (R) -- ( 0.50, -0.10);
        }%
    }}}%
}

\newcommand{\dblbox}{\mathord{\vcenter{\hbox{%
\tikz[
    baseline=-0.55ex,
    x=0.47em,
    y=0.47em,
    line cap=round,
    line join=round
]{
    \coordinate (n1) at (0,0.42);
    \coordinate (n2) at (0.72,0.42);
    \coordinate (n3) at (0.72,-0.42);
    \coordinate (n4) at (0,-0.42);
    \coordinate (n5) at (-0.42,0);

    \coordinate (g1) at (0.32,0.04);
    \coordinate (g2) at (0.40,-0.04);

    \coordinate (p1) at (-0.20,0.67);
    \coordinate (p2) at (0.95,0.67);
    \coordinate (p3) at (-0.20,-0.67);
    \coordinate (p4) at (-0.67,0);
    \coordinate (p5) at (0.95,-0.67);

    \draw[line width=0.42pt] (n1) -- (n2);
    \draw[line width=0.42pt] (n2) -- (n4);
    \draw[line width=0.42pt] (n4) -- (n5) -- (n1);
    \draw[line width=0.42pt] (n4) -- (n3);
    \draw[line width=0.42pt] (n1) -- (g1);
    \draw[line width=0.42pt] (g2) -- (n3);

    \draw[line width=0.85pt] (n1) -- (p1); 
    \draw[line width=0.42pt] (n2) -- (p2);
    \draw[line width=0.42pt] (n4) -- (p3);
    \draw[line width=0.42pt] (n5) -- (p4);
    \draw[line width=0.42pt] (n3) -- (p5);
}%
}}}}

\tcbuselibrary{breakable}

\newtcolorbox{simpleexample}[1][]{
  enhanced,
  breakable,
  colback=white,
  colframe=black!45,
  boxrule=0pt,
  borderline north={0.5pt}{0pt}{black!45},
  borderline south={0.5pt}{0pt}{black!45},
  left=0pt,
  right=0pt,
  top=6pt,
  bottom=6pt,
  before skip=12pt,
  after skip=12pt,
  #1
}

\definecolor{inactivecolor}{rgb}{0.2,0.2,0.2}
\definecolor{light-gray}{gray}{0.95}
\definecolor{gh-fg}{HTML}{000000}
\definecolor{gh-comment}{HTML}{007ACC}
\definecolor{gh-keyword}{HTML}{E60000}
\definecolor{gh-func}{HTML}{FF7F00}
\definecolor{gh-string}{HTML}{999999} 
\definecolor{gh-number}{HTML}{000000}
\definecolor{gh-linenum}{HTML}{666666}
\definecolor{gh-frame}{HTML}{BBBBBB}
\definecolor{disckosky-purple}{HTML}{22774d}
\definecolor{mred}{rgb}{0.70, 0.20, 0.20}
\definecolor{stringcolor}{gray}{0.5}
\newcommand{\githubrepo}{\href{https://github.com/Giu989/DiscKosky}{\faGithub}}

\usepackage{iftex}
\ifpdf\usepackage{pdfrender}\fi
\definecolor{disckosky-blue}{rgb}{0.1216,0.3725,0.7490}
\definecolor{disckosky-red}{rgb}{0.7843,0.0627,0.1804}
\makeatletter
\newdimen\DK@cap
\newdimen\DK@x
\newsavebox\DK@box
\newcommand\DK@hidden[2]{%
  \ifpdf\rlap{\resizebox{#1}{!}{\textpdfrender{TextRenderingMode=Invisible}{#2}}}\fi}
\DeclareRobustCommand\DiscKosky{%
  \leavevmode
  \begingroup
    \edef\DK@test{\noexpand\in@{b}{\f@series}}\DK@test
    \ifin@ \expandafter\DK@full \else \expandafter\DK@light \fi
  \endgroup}
\newcommand\DK@full{%
  \DK@cap=\fontcharht\font`D\relax
  \ifdim\DK@cap=\z@ \DK@cap=.7em\fi
  \DK@x=\fontcharht\font`x\relax
  \ifdim\DK@x=\z@ \DK@x=1ex\fi
  \DK@cap=\dimexpr(\DK@cap+\DK@x*783/591)/2\relax
  \sbox\DK@box{\raisebox{-\dimexpr\DK@cap*192/783\relax}{%
    \includegraphics[height=\dimexpr\DK@cap*975/783\relax]%
      {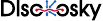}}}%
  \DK@hidden{\wd\DK@box}{DiscKosky}%
  \usebox\DK@box}
\newcommand\DK@light{%
  \DK@x=\fontcharht\font`x\relax
  \fontfamily{SourceCodePro-TLF}\fontseries{m}\fontshape{n}\selectfont
  \fontsize{\strip@pt\dimexpr\f@size pt*\DK@x/\fontcharht\font`x\relax}%
    {\f@baselineskip}\selectfont
  \pgfmathsetmacro\DK@em{\the\dimexpr1em\relax}%
  \pgfmathsetmacro\DK@cell{\the\fontcharwd\font`o/\DK@em}%
  \pgfmathsetmacro\DK@xc{-0.45*\DK@cell}%
  \def\DK@track{-0.05}%
  \pgfmathsetmacro\DK@box{\DK@cell+2*\DK@track}%
  \pgfmathsetmacro\DK@xo{\DK@box+0.5*\DK@cell}%
  \pgfmathsetmacro\DK@yw{0.5*\the\fontcharht\font`x/\DK@em}%
  \pgfmathsetmacro\DK@yt{\the\fontcharht\font`K/\DK@em-0.048}%
  \pgfmathsetmacro\DK@P{(\DK@xo-\DK@xc)/2.5}%
  \pgfmathsetmacro\DK@xs{\DK@xc+\DK@P}%
  \pgfmathsetmacro\DK@xr{\DK@xs+0.5*\DK@cell}%
  \def\DK@wave##1{\DK@yw+0.036*sin(360*(##1-\DK@xc)/\DK@P)}%
  \begin{tikzpicture}[baseline=0pt, x=1em, y=1em, inner sep=0pt, outer sep=0pt]
    \node[anchor=base west] (Disc) at (0,0)
      {D\kern\DK@track em i\kern\DK@track em s\kern\DK@track em c};
    \begin{scope}[shift={(Disc.base east)}]
      \node[anchor=base west] at (0,0) {\DK@hidden{\DK@box em}{K}};
      \begin{scope}[even odd rule]
        \clip (-1,-1) rectangle (4,1)
          plot[domain=\DK@xc:\DK@xo, samples=41, variable=\t]
            (\t, {\DK@wave{\t}+0.043})
          -- plot[domain=\DK@xo:\DK@xc, samples=41, variable=\t]
            (\t, {\DK@wave{\t}-0.043}) -- cycle;
        \node[anchor=base west] (osky) at (\DK@box,0)
          {o\kern\DK@track em s\kern\DK@track em k\kern\DK@track em y};
      \end{scope}
      \begin{scope}[disckosky-blue, line width=0.05em, line cap=round,
                    line join=round]
        \draw (\DK@xs,0.048) -- (\DK@xs,\DK@yt)
          (\DK@xr,\DK@yt) -- (\DK@xs,\DK@yw) -- (\DK@xr,0.048);
        \fill (\DK@xs,0.048) circle[radius=0.048]
          (\DK@xs,\DK@yt) circle[radius=0.048]
          (\DK@xr,0.048) circle[radius=0.048]
          (\DK@xr,\DK@yt) circle[radius=0.048];
      \end{scope}
      \begin{scope}[disckosky-red]
        \draw[line width=0.05em, line cap=round]
          plot[domain=\DK@xc:\DK@xo, samples=61, smooth, variable=\t]
            (\t, {\DK@wave{\t}});
        \fill (\DK@xc,\DK@yw) circle[radius=0.058]
          (\DK@xo,\DK@yw) circle[radius=0.058];
      \end{scope}
    \end{scope}
    \pgfresetboundingbox
    \useasboundingbox (Disc.base west |- osky.south) rectangle (osky.north east);
  \end{tikzpicture}}
\makeatother

\definecolor{charcoal}{HTML}{343837}

\title{
    \centering
    \includegraphics[width=0.4\textwidth]{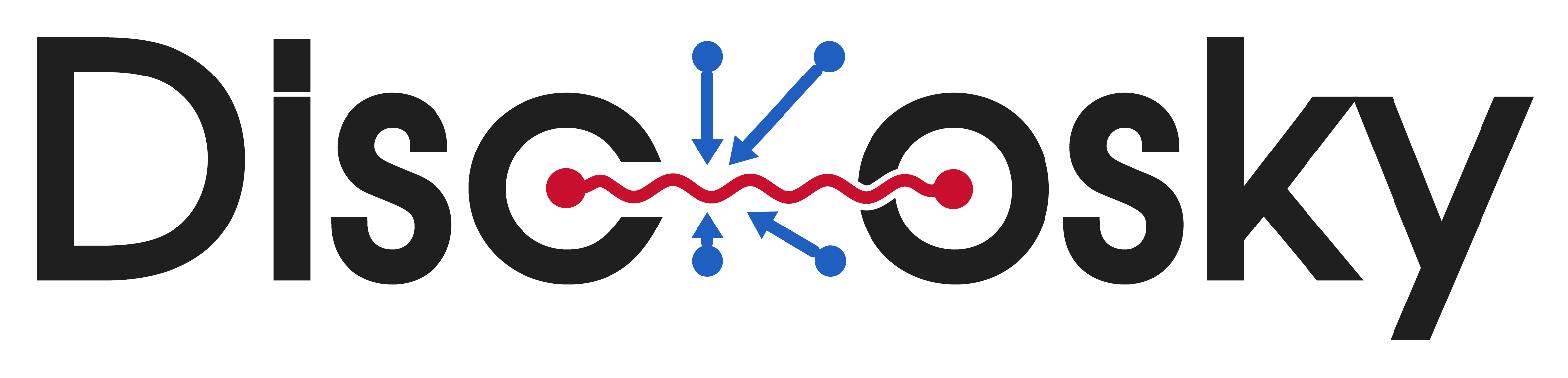}\\
    Efficient Genealogical Constraints from Sector-By-Sector Critical Point Counting
}
\author{Giulio~Crisanti, Luke~Lippstreu, Andrew~J.~McLeod, and Maria~Polackova}
\affiliation{Higgs Centre for Theoretical Physics, School of Physics and Astronomy,
The University of Edinburgh, Edinburgh EH9 3FD, Scotland, UK}

\abstract{
We present \DiscKosky{}, a new \textsc{Mathematica} package for efficiently generating genealogical constraints on the sequential discontinuities of Feynman integrals. \DiscKosky{} works by tracking the solutions to the critical point equations associated with a Feynman integral; by identifying which of these critical points cease to exist in a given kinematic limit, we can associate each Landau singularity with a set of on-shell sectors. The hierarchical principle then forbids sequential discontinuities with respect to singularities that are associated with incompatible sets of on-shell propagators. In rare cases, this algorithm fails to make predictions due to the appearance of higher-dimensional critical varieties---nevertheless, we show that this phenomenon can sometimes be used to identify non-repeating discontinuities, which can be computed at most once in a Feynman integral. We illustrate the power of \DiscKosky{} by applying the package to a broad class of Feynman integrals, including two-loop five-point integrals with one and two massive legs, and three-loop examples that go beyond the current computational frontier. \DiscKosky{} can also be used to probe the analytic properties of more general classes of Euler integrals, which we illustrate by considering several examples that arise in the context of cosmological correlators.
}

\begin{document}

\maketitle

\include{sections/Introduction}

\include{sections/Feynman_Integrals_and_Landau_Analysis}

\include{sections/quick_starter_guide}
\include{sections/Examples}

\include{sections/non_repeating_letters}

\include{sections/genelogical_on_individual_symbol}

\include{sections/Conclusions}

\section*{Acknowledgments}

We thank Ben Page and Erik Panzer for insightful discussions.  The research of GC is supported by the United Kingdom Research and Innovation
grant UKRI FLF MR/Y003829/1, while AJM and LL acknowledge support from the Royal Society grant
URF{\textbackslash}R1{\textbackslash}221233. AJM additionally acknowledges support from the European Research Council (ERC) under the European Union’s Horizon Europe research and innovation program grant agreement 101163627 (ERC Starting Grant ``AmpBoot'').

\bibliographystyle{jhep}
\bibliography{refs}
\end{document}

%% file: sections/Introduction.tex
\section{Introduction}
The evaluation of Feynman integrals remains a major computational bottleneck across many areas of physics, from precision collider phenomenology and gravitational-wave physics to the computation of cosmological correlators. This challenge has motivated the development of new methods for probing the mathematical structure of these integrals, drawing on ideas from algebraic geometry, topology, number theory, combinatorics, complex analysis, and the theory of differential equations. Within this broader effort, particular attention has been paid to the analytic properties of scattering amplitudes (and related quantities in quantum field theory), and the question of how these analytic properties are constrained by principles such as causality and unitarity. A central aspect of this question concerns where Feynman integrals can become singular, and what sequences of discontinuities they encode.

Feynman integrals become singular in kinematic configurations where the on-shell hypersurfaces that appear in the integrand pinch the integration contour. 
By analytically continuing around these singular loci in the space of external kinematics, one can probe the branch cut structure of a given integral, uncovering important information about the spaces of special functions it evaluates to.
For instance, much use has been made of the Steinmann relations~\cite{Steinmann,Steinmann2,Cahill:1973qp}, which state that double discontinuities computed with respect to pairs of partially overlapping momentum channels $s$ and $t$ must vanish:
\begin{gather}
    \text{Disc}_{t-M_t^2}\, \text{Disc}_{s-M_s^2} \, \mathcal{I} = 0  \, ,\label{eq:steinmann}
\end{gather}
where $M_s$ and $M_t$ represent the sums of masses that cross the $s$ and $t$ cuts\footnote{For instance, for the box integral with equal internal masses $m$, we would have $s=(p_1+p_2)^2$, $t=(p_2+p_3)^2$, and $M_s = M_t = 2m$. Note that the Steinmann relations have only been proven to hold for certain classes of amplitudes and Feynman integrals~\cite{Cahill:1973qp,Bourjaily:2020wvq}. However, they have been observed to hold much more broadly; for a more extensive discussion, see~\cite{Hannesdottir:2025bss}.}. In this expression, each discontinuity is to be understood as the difference between the integral before and after it has been analytically continued around the corresponding branch point---in other words, we can schematically write the discontinuity operator as $\text{Disc}_\lambda = \left(\circlearrowleft_\lambda - 1 \right)$.
The Steinmann relations follow from causality, and have played a central role in bootstrapping amplitudes in the planar limit of maximally supersymmetric Yang-Mills theory to high loop orders~\cite{Caron-Huot:2016owq,Caron-Huot:2020bkp,Dixon:2022rse,Arkani-Hamed:2022rwr,Dixon:2023kop,Basso:2024hlx}. The surprising power of these relations has motivated a search for more extensive analytic constraints that can be placed on amplitudes and Feynman integrals in more general quantum field theories~\cite{Drummond:2017ssj,Caron-Huot:2019bsq,Benincasa:2020aoj,Hannesdottir:2021kpd,Hannesdottir:2024cnn,Bargiela:2026lje}, such as the Standard Model.

One promising avenue for uncovering further constraints is via the \textit{hierarchical principle}~\cite{pham,Landshoff1966,boyling1968homological}. This principle leverages the idea that each of a Feynman integral's singularities is associated with a collection of on-shell hypersurfaces (as read off the corresponding solution to the Landau equations~\cite{landau1959}), and that computing a discontinuity with respect to a given singularity localizes the integration contour to this set of hypersurfaces. At the same time, not all of a Feynman integral's singularities can be kinematically reached once certain propagators are on-shell. This implies that some sequential discontinuities must vanish; namely,
\begin{equation}
    \text{Disc}_{\lambda^\prime}\, \cdots \, \text{Disc}_{\lambda} \, \cdots \,\mathcal{I} = 0 \qquad \text{ if  } \qquad S_\lambda \not \subseteq S_{\lambda^\prime}
    \label{eq:hierarchical-principle-intro}
\end{equation}
where $S_{\lambda}$ represents the set of propagators whose on-shell surfaces are associated with the singularity at $\lambda=0$.\footnote{In general, there can be multiple sets of on-shell surfaces that give rise to singularities at $\lambda=0$ or $\lambda^\prime=0$; all such singularities must be taken into account when deriving constraints of the form~\eqref{eq:hierarchical-principle-intro}.} Here, the ellipses indicate that the same relation holds even if an arbitrary sequence of discontinuities or analytic continuations is carried out before applying each discontinuity operator. Moreover, these constraints hold for arbitrary spacetime dimension, so they are not spoiled by infrared divergences and apply to all orders in dimensional regularization.

While it has long been suspected that~\eqref{eq:hierarchical-principle-intro} places severe constraints on the discontinuity structure of individual Feynman integrals, extracting concrete predictions in specific examples has long proven computationally intractable (due to the fact that deriving these constraints in full detail requires carrying out complicated algebraic blowups; see for instance~\cite{Berghoff:2022mqu}). This has limited the hierarchical principle's practical utility for most of the last six decades. However, recent conceptual and computational advances have made predictions of the form~\eqref{eq:hierarchical-principle-intro} considerably easier to derive. A relative version of the Picard--Lefschetz formula that applies to Feynman integrals in parametric form was established in~\cite{Berghoff:2022mqu}, and several explicit examples have been analyzed by direct analytic continuation~\cite{Britto:2023rig,Britto:2025wzt}. In parallel, new topological methods for detecting the singularities of Feynman integrals were developed in~\cite{Fevola:2023kaw,Fevola:2023fzn}. Building on these advances, a new, computationally tractable algorithm was devised in~\cite{Hannesdottir:2024cnn} for deriving \emph{genealogical constraints}, which can be thought of as those hierarchical constraints that can be detected by probing (the change in) efficient-to-compute topological invariants.

In more detail, the idea underlying genealogical constraints is that one can determine which singularities can be reached within a given on-shell sector by probing the number of solutions to the unregulated critical-point equations~\cite{Lee_2013}, namely
\begin{gather}
    \begin{gathered}
        \partial_{\alpha_e}\mathcal{G}=0
        \quad \forall\, e\in S,\\
        \alpha_e=0
        \quad \forall\, e\notin S,
    \end{gathered}
    \qquad
    \text{with}\quad
    \mathcal{G}\neq 0,\infty ,
    \label{eq:crit pt eqn}
\end{gather}
where $\mathcal{G}$ is the Lee--Pomeransky polynomial, and $S$ lists which propagators are on-shell in the current sector. In particular, on kinematic hypersurfaces where a given Feynman integral becomes singular, the number of isolated solutions to these equations must decrease in one or more sectors. By checking which sectors this change occurs in, one can infer which propagators must go on shell in order to access the corresponding singularity.

Once these sectors have been identified, the hierarchical principle can be applied directly by comparing the sectors associated with different singularities, and applying~\eqref{eq:hierarchical-principle-intro}. Note that this approach makes use of more refined information than minimal cuts, which were introduced in~\cite{Hannesdottir:2024cnn} as an easy graphical way to infer the minimal sets of propagators that must be involved in pinching the integration contour. In practice, minimal cuts still provide a reliable lower bound on which propagators will be on shell after a given discontinuity has been computed, which can prove useful in cases where the critical point equations~\eqref{eq:crit pt eqn} must be regulated. 

In~\cite{Hannesdottir:2024cnn}, it was found (in a number of two-loop examples) that nearly all of the implications of the hierarchical principle could be deduced as genealogical constraints using minimal cuts. In this paper, we automate this type of genealogical analysis, while also extending its reach through the use of sector-by-sector critical point counting and through the identification of new non-repeating discontinuity constraints (which amount to statements of the form~\eqref{eq:hierarchical-principle-intro} in which $\lambda=\lambda^\prime$). These non-repeating discontinuities arise when a singularity is associated with a sector in which there are no critical points in generic kinematics; when one computes a discontinuity with respect to such a singularity, one ends up in (some linear combination of) sectors in which additional propagators are on shell, implying the same singularity can no longer be accessed~\cite{Bargiela:2026lje}. With these extensions, we find that we can deduce all of the implications of the hierarchical principle in a number of nontrivial examples. 

There remain edge cases in which these methods do not immediately apply, due to subtleties involving higher-dimensional solutions to the critical point equations~\eqref{eq:crit pt eqn}. Even so, these edge cases can be easily detected, and the number of solutions to the regulated critical point equations can be computed as a cross-check on the unregulated sector-by-sector analysis, ensuring that no incorrect predictions are made.

The algorithms presented in this paper are implemented in a new \textsc{Mathematica} package \DiscKosky{}, which generates the set of genealogical constraints that can be derived using the above-described methods, for any given Feynman integral. As input, the package requires the Lee--Pomeransky polynomial and a list of candidate singular loci. When the critical-point analysis described above is insufficient to draw a reliable conclusion, \DiscKosky{} instead returns an indeterminate result (which can be conservatively interpreted as the sequence of discontinuities being allowed). However, some of these indeterminate results can be refined into non-repeating discontinuity constraints, further strengthening the predictive power of the algorithm. 

We have tested the efficiency and efficacy of this code on a broad range of two- and three-loop Feynman integrals; in these examples, \DiscKosky{} predicts thousands of vanishing double discontinuities, reproducing all of the hierarchical restrictions associated with logarithmic branch points in the two-loop families considered below, and nearly all of the additional restrictions that arise once algebraic branch points are included. \DiscKosky{} can also be used to generate analogous constraints on more general Euler integrals, which we illustrate by considering several integrals that appear in the context of cosmological correlators.

The remainder of this paper is organized as follows. In Section~\ref{sect:theoretical-background}, we review the Landau equations and the hierarchical principle, and explain in detail how sector-by-sector critical-point counting can be used to determine the on-shell information needed to derive genealogical constraints. Section~\ref{sec:quickstart} then provides a practical introduction to \DiscKosky{}, including installation instructions and an overview of its main functions. In Section~\ref{sec:examples}, we illustrate the scope of the package on a range of nontrivial two- and three-loop Feynman integrals, as well as more general Euler integrals. Then, in Section~\ref{sec:non_repeating}, we discuss non-repeating discontinuities, and show how changes in the dimension of the critical locus provide new, previously unleveraged constraints. Finally, in Section~\ref{sec:algebraic-letters} we illustrate how these discontinuity constraints work in examples that involve both algebraic and logarithmic branch points, before concluding in Section~\ref{sec:conclusions}. The \DiscKosky{} code can be found at the following GitHub repository: \githubrepo.

%% file: sections/Feynman_Integrals_and_Landau_Analysis.tex
\section{Theoretical Background}\label{sect:theoretical-background}
In this section, we review all of the background material needed to explain genealogical constraints and the methods implemented in \DiscKosky{}. First, we review the Landau equations and how the singularities of Feynman integrals are associated with specific on-shell sectors. We then explain how this understanding naturally gives rise to the hierarchical principle---first articulated in the sixties~\cite{pham,Landshoff1966,boyling1968homological}---which restricts the sequences of discontinuities that can appear in a given integral. Building on this perspective, we show how hierarchical constraints can be derived by tracking solutions to the critical-point equations, providing a practical alternative to solving the Landau equations directly. We then discuss subtleties that can arise in critical-point counting, and explain how regulated critical-point equations provide a consistency check on our methods. Finally, we describe how these counts can be performed efficiently using Gr\"obner bases over finite fields. We end with a summary of the key background from this section that is needed to follow the rest of the paper.

\subsection{The Landau Equations}\label{sect:landau-equations}
In momentum space, Feynman integrals take the form
\begin{equation}
	I_{\vec{\nu}}(\vec{s})
	=
	\int
	\frac{\mathrm{d}^{D}\ell_{1}\,\mathrm{d}^{D}\ell_{2}\cdots \mathrm{d}^{D}\ell_{L}}
	{\left(q_{1}^{2}-m_{1}^{2}\right)^{\nu_{1}}
		\left(q_{2}^{2}-m_{2}^{2}\right)^{\nu_{2}}
		\cdots
		\left(q_{N}^{2}-m_{N}^{2}\right)^{\nu_{N}}}
	\,,\label{eq:gen feyn}
\end{equation}
where $D$ and $L$ are the spacetime dimension and the number of loops, $\ell_i$ denotes the $i^\text{th}$ loop momentum, and $q_e$ and $m_e$ are
the momentum and mass of internal edge $e$. We assume the propagator powers $\nu_e$ are all integers and collect them in the vector $\vec{\nu}$, while similarly denoting the collection of external kinematic variables (which include both Mandelstam invariants and internal masses $m_e$) by $\vec{s}$.\footnote{Note that the restriction to integer powers is important in this context---if the $\nu_i$ powers are allowed to take general complex values, then we cannot probe the critical points of these integrals using the unregulated critical point equations.}

The kinematic hypersurfaces on which Feynman integrals can become singular are characterized by the Landau equations \cite{10.1143/PTP.22.128,Landau:1959fi,Bjorken:1959fd}. Starting from~\eqref{eq:gen feyn}, one asks where in the space of external kinematics $\vec{s}$ the loop-momentum integration contour can become pinched by the on-shell surfaces $q_e^2 - m_e^2 = 0$. A necessary condition for such a pinch to occur is that there exists a collection of parameters $\alpha_e$ satisfying
\begin{align}
	\alpha_e(q_e^2-m_e^2)=0,\qquad &e=1,2,\ldots,N\, ,\label{eq:landau-on-shell}\\
	\frac{\partial }{\partial \ell_j^{\mu}}\sum_{e=1}^{N}\alpha_e(q_e^2-m_e^2)=0,\qquad &j=1,2,\ldots,L\, ,\label{eq:landau-stationary}
\end{align}
in which at least one of the $\alpha_e$ parameters is nonzero. The first of these sets of equations determines which propagators are involved in pinching the integration contour---namely, for each internal line, either $\alpha_e=0$ (indicating propagator $e$ is not involved) or the corresponding propagator is placed on shell, namely $q_e^2-m_e^2=0$.\footnote{It is also possible for both factors in~\eqref{eq:landau-on-shell} to vanish; for a discussion of such cases see~\cite{Hannesdottir:2022xki}.} The second set imposes the constraint that these on-shell surfaces intersect non-transversely in the space of loop momenta.\footnote{Note that a single on-shell surface can satisfy this condition by itself if it becomes singular, namely if its normal vector vanishes. For instance, for a massless propagator, this can happen at $q_e^\mu=0$.} Thus, we see that each solution of the Landau equations not only identifies a possible singular locus of the integral, but also the set of on-shell hypersurfaces responsible for it.

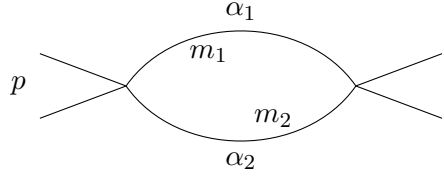
\begin{figure}[t]
    \centering
    \begin{tikzpicture}[scale=0.95]
        \coordinate (L) at (-1.6,0);
        \coordinate (R) at ( 1.6,0);

        \draw (L) -- (-2.8,  0.45);
        \draw (L) -- (-2.8, -0.45);
        \draw (R) -- ( 2.8,  0.45);
        \draw (R) -- ( 2.8, -0.45);

        \draw (L) to[out=55,in=125] (R);
        \draw (L) to[out=-55,in=-125] (R);

        \node at (0,1.04) {$\alpha_1$};
        \node at (0,-1.04) {$\alpha_2$};

        \node at (-0.45,0.45) {$m_1$};
        \node at ( 0.45,-0.45) {$m_2$};
        \node at (-3.1,0) {$p$};
    \end{tikzpicture}
    \caption{The two-mass bubble integral with four massless external legs.}
    \label{fig:bubble}
\end{figure}

As an example, consider the bubble integral with different internal masses, as shown in Figure~\ref{fig:bubble}:
\begin{equation}\label{eq:bubble}
	\mathcal{I}_{\hspace{.01cm}\FeynmanBubble\hspace{.01cm}}(p^2,m_1^2,m_2^2)
	=
	\int
	\frac{\mathrm{d}^{D}\ell}
	{\left(\ell^{2}-m_{1}^{2}\right)
		\left((\ell+p)^{2}-m_{2}^{2}\right)}\, .
\end{equation}
This integral has a singularity at $m_1=0$, which arises from a solution to the Landau equations in which only the propagator with mass $m_1$ is on shell. Likewise, a singularity at $m_2=0$ can be reached when just the propagator with mass $m_2$ is on shell. Moreover, singularities at these kinematic loci no longer arise if we place both propagators on shell. Conversely, the threshold and pseudo-threshold singularities at $p^2=(m_1\pm m_2)^2$ can \emph{only} arise if both propagators are on shell. Outside of $D=2$, there is also a second-type singularity at $p^2=0$, which requires both propagators to be on shell. (For a more detailed discussion and derivation of these solutions, see for instance~\cite{Hannesdottir:2022xki}.) This information is collected in Table~\ref{tab:bubble-onshell-lines}, where we label each sector by the set of edges that are on shell.

\begin{table}[b]
	\centering
	\begin{tabular}{c|c}
		\textbf{Singularity} & \textbf{On-Shell Lines} \\
		\hline
		$m_1=0$ & $\{1\}$ \\
		$m_2=0$ & $\{2\}$ \\
		$p^2=(m_1\pm m_2)^2$ & $\{1,2\}$ \\
		$p^2=0$ & $\{1,2\}$
	\end{tabular}
	\caption{Singularities of the two-mass bubble integral and their corresponding on-shell sectors.}
	\label{tab:bubble-onshell-lines}
\end{table}

\begin{figure}[t]
	\centering
	\raisebox{-1.6cm}{\begin{tikzpicture}[scale=1.2]
			\coordinate (a) at (-.8,.8);
			\coordinate (b) at (.8,.8);
			\coordinate (c) at (.8,-.8);
			\coordinate (d) at (-.8,-.8);
			
			\node (a1) at (-1.5, 1.5) {$p_2$};
			\node (d1) at (-1.5, -1.5) {$p_1$};
			\node (b1) at (1.5, 1.5) {$p_3$};
			\node (c1) at (1.5, -1.5) {$p_4$};
			
			\draw[] (a) to node[midway, above] {$\alpha_2$} (b)
			to node[midway, right] {$\alpha_3$} (c)
			to node[midway, below] {$\alpha_4$} (d)
			to node[midway, left] {$\alpha_1$} (a);
			
			\draw[] (a1) to (a);
			\draw[] (d1) to (d);
			\draw[] (b1) to (b);
			\draw[] (c1) to (c);
	\end{tikzpicture}}
	\caption{The massless box integral.}
	\label{fig:box}
\end{figure}
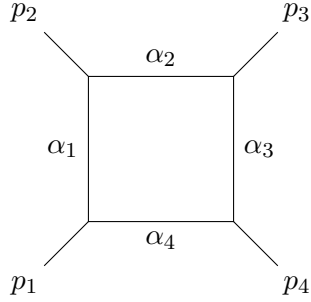

In more complicated examples, singularities can arise on a given kinematic hypersurface $\lambda(\vec{s})=0$ in multiple ways. This can already be seen in the massless box integral,
\begin{equation} 
	I_{\masslessbox}(s_{12},s_{23})
	=
	\int
	\frac{\mathrm{d}^{D}\ell}
	{\ell^{2}\,(\ell+p_{1})^{2}\,(\ell+p_{1}+p_{2})^{2}\,(\ell-p_{4})^{2}}
	\, ,
	\label{eq:massless_box_integral}
\end{equation}
where a singularity can arise at $s_{23}=(p_2+p_3)^2=0$ in configurations involving two or four on-shell propagators, as summarized in Table~\ref{tab:massless-box-landau-solutions}. For more discussion of the corresponding solutions to the Landau equations for the massless box, see~\cite{Hannesdottir:2025bss}.

\begin{table}[b]
	\centering
	\begin{tabular}{c|c}
		\textbf{Singularity} & \textbf{On-Shell Sectors} \\
		\hline \\[-.38cm]
		$s_{23}=(p_2+p_3)^2=0$
		&
		$\{1,3\}$\,,
		$\{1,2,3,4 \}$
		\\[3pt]
		$s_{12}=(p_1+p_2)^2=0$
		&
		$\{2,4\}$\,,
		$\{1,2,3,4\}$\\[3pt]
		$s_{12}+s_{23}=0$
		&
		$\{1,2,3,4\}$
	\end{tabular}
	\caption{Landau solutions for the massless box integral. For each kinematic singularity, the table lists the sets of on-shell lines that can give rise to a singularity on the corresponding kinematic locus.}
	\label{tab:massless-box-landau-solutions}
\end{table}

\subsection{The Hierarchical Principle}\label{sect:hierarchical}

The on-shell information captured by the Landau equations has direct consequences for the discontinuity structure of Feynman integrals. The hierarchical principle manifests one important facet of this structure, namely the fact that computing discontinuities necessarily places propagators on shell~\cite{pham2011singularities,Berghoff:2022mqu}. More specifically, in momentum space, computing $\text{Disc}_\lambda$ modifies the integration contour in such a way that all of the propagators in $S_\lambda$ are now fixed to be on shell, where $S_\lambda$ denotes the set of propagators that are required to satisfy $q_e^2-m_e^2=0$ in the corresponding solution to the Landau equations. As a result, subsequent discontinuities must all be accessible on the support of these on-shell conditions.

When a singularity \(\lambda=0\) admits multiple solutions to the Landau equations, there may be multiple collections of on-shell surfaces that can give rise to a nonzero discontinuity. In such cases, we write $S_{\lambda,i}$ to denote the set of on-shell propagators associated with the $i^\text{th}$ such solution. In this language, the hierarchical principle implies the following class of constraints on the sequential discontinuities that can appear in the corresponding Feynman integral:
\begin{tcolorbox}[
	enhanced,
	arc=2mm,
	boxrule=0.6pt,
	left=2mm,right=2mm,top=1.2mm,bottom=1.2mm,
	colback=white,
	colframe=black,
	]
	\textbf{The Hierarchical Principle:}
\begin{equation}
	\left(
	\forall\, i,j:\;
	S_{\lambda',j}\not\supseteq S_{\lambda,i}
	\right)
	\quad\Longrightarrow\quad
	\text{Disc}_{\lambda'}\,\cdots\,
	\text{Disc}_{\lambda}\,\cdots\,\mathcal{I}
	=0\, .
	\label{eq:hierarchical-principle-background}
\end{equation}
\end{tcolorbox}
In words, \eqref{eq:hierarchical-principle-background} says that  $\text{Disc}_{\lambda'}$ can return a nonzero result only if the singularity at $\lambda^\prime=0$ can still be kinematically reached after $\text{Disc}_{\lambda}$ has been computed. If not, all sequences of discontinuities taking the form~\eqref{eq:hierarchical-principle-background} must vanish. 

Note that, for first discontinuities, the hierarchical principle is naturally encoded in Cutkosky's formula, which replaces all of the propagators that are cut in a given momentum channel by delta functions~\cite{Cutkosky:1960sp}. Clearly, the only discontinuities that can still arise after this replacement are those that can still be reached on the support of the corresponding on-shell constraints. The hierarchical principle extends this idea to all sequences of discontinuities, even though the standard version of Cutkosky's formula no longer holds.

A slightly different picture holds in Feynman-parameter space. Recall that in the Lee--Pomeransky representation, Feynman integrals take the form
\begin{equation}
	I_{\vec{\nu}}(\vec{s})
	\propto
	\int_{\alpha_e\ge 0}
	\mathrm{d}^N\alpha\,
	\frac{\prod_{e=1}^{N}\alpha_e^{\nu_e-1}}
	{\mathcal{G}(\vec{\alpha},\vec{s})^{D/2}}\, ,
	\label{eq:lee-pomeransky-representation}
\end{equation}
where the polynomial
\begin{gather}
	\mathcal{G}(\alpha;\vec{s})=\mathcal{U}+\mathcal{F}
	\label{eq:lee-pomeransky-graph-polynomial}
\end{gather}
is the sum of the first and second Symanzik polynomials $\mathcal{U}$ and $\mathcal{F}$.\footnote{Note that the $\alpha_e$ variables in this representation can be identified with the ones appearing in the Landau equations presented in the previous subsection; to see this, one can rederive the Landau equations after Feynman parameters have been introduced, but before the loop momentum has been integrated out---doing so, one arrives back at~\eqref{eq:landau-on-shell} and~\eqref{eq:landau-stationary}.} Since, in this representation, the integrand involves only a single polynomial $\mathcal{G}$ with dependence on external kinematics, every kinematic discontinuity must localize the integration contour to the $\mathcal{G}=0$ locus. However, unlike in the loop momentum representation, we now have integration boundaries---and this is where the hierarchical principle appears. Namely, the integration contour that one is left with after computing a discontinuity in Feynman parameter space will no longer depend on the $\alpha_e=0$ endpoints associated with the propagators that the corresponding solution to the Landau equations puts on shell~\cite{Berghoff:2022mqu,Britto:2023rig}. Subsequent discontinuities can further deform the new contour, but can never restore one of these integration boundaries once it has been lost. Thus, as one computes additional discontinuities, the set of $\alpha_e=0$ integration boundaries naturally gets smaller, mirroring the set of propagators that are forced to remain on shell in loop-momentum space. 

The consequences of the hierarchical principle can already be seen in the two examples from the previous subsection. For the two-mass bubble~\eqref{eq:bubble}, Table~\ref{tab:bubble-onshell-lines} shows that the singularities at $m_1=0$ and $m_2=0$ are associated with the on-shell sets $\{1\}$ and $\{2\}$, respectively. Since neither set contains the other, computing a discontinuity around either mass singularity prevents a subsequent discontinuity around the other:
\begin{equation}
	\text{Disc}_{m_i}\,\cdots\,
	\text{Disc}_{m_j}\,\cdots\,
	\mathcal{I}_{\hspace{.01cm}\FeynmanBubble\hspace{.01cm}}	=0\, ,
	\qquad i\neq j\, .
\end{equation}
Similarly, the threshold and pseudo-threshold singularities at $p^2=(m_1\pm m_2)^2$, as well as the second-type singularity at $p^2=0$, are associated with the sector $\{1,2\}$. Once any of these discontinuities has been computed, the hierarchical principle therefore forbids a discontinuity from appearing at either $m_1=0$ or $m_2=0$:
\begin{align}
	\text{Disc}_{m_i}\,\cdots\,
	\text{Disc}_{p^2-(m_1\pm m_2)^2}\,\cdots\,
	\mathcal{I}_{\hspace{.01cm}\FeynmanBubble\hspace{.01cm}}
	&=0\, ,
	\qquad i=1,2\, ,\\
	\text{Disc}_{m_i}\,\cdots\,
	\text{Disc}_{p^2}\,\cdots\,
	\mathcal{I}_{\hspace{.01cm}\FeynmanBubble\hspace{.01cm}}
	&=0\, ,
	\qquad i=1,2\, .
\end{align}
These relations can be verified to all orders in dimensional regularization (for instance, using the closed-form expression in~\cite{Abreu:2017mtm}).

The massless box provides an instructive contrast. In this case, the hierarchical principle does \emph{not} tell us that sequential discontinuities of the form
\begin{equation}
	\text{Disc}_{s_{12}}\,\cdots\,
	\text{Disc}_{s_{23}}\,\cdots\,
	\mathcal{I}_{\masslessbox}
\end{equation}
must vanish. As shown in Table~\ref{tab:massless-box-landau-solutions}, the singularity at $s_{12}=0$ admits a Landau solution on the maximal cut, where all four propagators are on shell. The corresponding singularity can thus be reached after one has computed a discontinuity with respect to either of the singularities at $s_{23}=0$. Note that this example also violates the (na\"ive version of the) Steinmann relations due to these maximal-cut singularities; for a more complete discussion, see~\cite{Hannesdottir:2025bss}.

As illustrated in these examples, once the sets $S_{\lambda,i}$ associated with each singularity are known, applying the hierarchical principle just requires comparing these sets by inclusion. The main challenge is therefore determining which collections of on-shell surfaces can pinch the contour to give rise to a singularity on a given kinematic hypersurface. As we now explain, the Feynman-parameter interpretation developed above provides a way to extract precisely this information by studying how the solutions to the critical-point equations change in each sector.

\subsection{The Critical Point Equations}\label{sect:critical-point-counting}

A central insight from Picard--Lefschetz theory is that the singularities of integrals are reflected in how the space of possible integration cycles changes as external parameters are varied.\footnote{Picard--Lefschetz theory has various applications in physics, including the analytic continuation of Chern--Simons theory \cite{witten2010analyticcontinuationchernsimonstheory}, Liouville theory \cite{Harlow_2011}, lattice QCD \cite{Cristoforetti:2012su}, the wavefunction of the Universe \cite{Feldbrugge_2017}, and tree-level scattering in string theory \cite{Mizera_2020}. See in particular the discussion in~\cite{Hannesdottir:2022bmo}, which highlights the relationship between Lefschetz thimbles and singularities of Feynman integrals.} In particular, a nontrivial discontinuity around a locus $\lambda=0$ can arise only if there exists an integration cycle that vanishes in the strict $\lambda \to 0$ limit~\cite{pham,Berghoff:2022mqu}. If the number of independent cycles remains unchanged when $\lambda=0$, then the operator $\text{Disc}_\lambda$ must return zero.

This observation turns the problem of identifying nonzero discontinuities into a counting problem. And in fact, Picard--Lefschetz theory provides a practical way to perform this count. The set of independent integration cycles forms a basis for the relevant homology group, whose dimension can (in suitably generic situations) be computed from the Euler characteristic of the integration space. The same Euler characteristic can in turn be obtained by counting the number of isolated critical points~\cite{Huh_2013}. Thus, generically, the number of independent integration cycles is equal to the number of critical points of $\mathcal{G}$, which can be found as the solutions to the following equations:
\begin{tcolorbox}[
    enhanced,
    arc=2mm,
    boxrule=0.6pt,
    left=2mm,
    right=2mm,
    top=1.2mm,
    bottom=1.2mm,
    colback=white,
    colframe=black,
]
    \textbf{Critical Point Equations:}
\begin{equation}
    \begin{gathered}
        \alpha_e\,\partial_{\alpha_e}\mathcal{G}=0,
        \qquad e=1,\ldots,N, \\
        \mathcal{G}(\vec{\alpha},\vec{s})\neq 0
    \end{gathered}
    \label{eq:critical-point-equations}
\end{equation}
\end{tcolorbox}
These equations differ from the Landau equations~\eqref{eq:landau-on-shell} and~\eqref{eq:landau-stationary} insofar as the latter require that $\mathcal{G} = 0,\infty$~\cite{Fevola:2023fzn,Fevola:2023kaw} when formulated in the Lee--Pomeransky representation. However, similar to the Landau equations, the solutions to~\eqref{eq:critical-point-equations} are naturally organized into sectors corresponding to the coordinate strata of Feynman-parameter space.\footnote{Note that the critical point equations are solved in affine space $\mathbb{C}^N$, so solutions at infinity do not form part of the solution set.} In particular, the sector of a critical point $\sigma$ is defined to be
\begin{equation}
	S\big(\sigma\big) = \big\{\, e : \partial_{\alpha_e}\mathcal{G}\big(\sigma\big)=0 \,\big\}\,. \label{eq:critical_point_sector}
\end{equation}
We also denote by $N_S$ the number of isolated critical points in sector $S$, counted with multiplicity. Keeping track of the sector in which each critical point arises is essential, since this is what allows us to determine the on-shell sectors to associate with each singularity. 

In generic kinematics, each isolated critical point can be associated with an independent integration cycle (namely, a Lefschetz thimble). After setting $\lambda=0$, some of these critical points may cease to exist; the associated integration cycles, which degenerate in the strict $\lambda\to0$ limit, constitute vanishing cycles. Thus, a decrease in the number of critical points in a given sector signals the existence of a vanishing cycle in that sector. By comparing the critical-point count sector by sector at generic kinematics and on the locus $\lambda=0$, we can therefore determine which sectors are associated with the corresponding singular hypersurface.

Although we have identified these sectors in Feynman parameter space, they are in one-to-one correspondence with the on-shell sectors discussed in loop momentum space. As explained above, a cycle in sector $S$ no longer depends on the boundaries $\alpha_e=0$ for $e\in S$. By~\eqref{eq:landau-on-shell}, if the Landau equation for edge \(e\) cannot be satisfied by fixing $\alpha_e$ to the endpoint at \(0\), the corresponding propagator must be on shell. We can thus identify each sector in which the critical-point count decreases with one of the on-shell sets $S_{\lambda,i}$. This leads to the following practical criterion for determining the on-shell configurations associated with a given singularity:
\begin{tcolorbox}[
    enhanced,
    arc=2mm,
    boxrule=0.6pt,
    left=2mm,
    right=2mm,
    top=1.2mm,
    bottom=1.2mm,
    colback=white,
    colframe=black,
]
    \phantomsection
    \label{crit:critical-point-counting}
    \textbf{Critical-Point-Counting Criterion:}

    For a candidate singularity $\lambda(\vec{s})=0$:
    \begin{enumerate}
        \item In generic kinematics, count the number of isolated solutions to~\eqref{eq:critical-point-equations} in each sector.
        \item Repeat the count after restricting the external kinematics to the locus $\lambda(\vec{s})=0$.
        \item Each sector $S$ for which $N_S$ decreases after restricting to
$\lambda(\vec{s})=0$ identifies an on-shell set $S_{\lambda,i}$
associated with the singularity $\lambda(\vec{s})=0$.
    \end{enumerate}
\end{tcolorbox}
Crucially, this method only requires us to detect a decrease in the number of critical points; we do not need to determine how the individual critical points are lost.

Let us illustrate the critical-point-counting criterion using the two-mass bubble from~\eqref{eq:bubble}. The Lee--Pomeransky polynomial for this diagram is
\begin{equation}
    \mathcal{G}_{\hspace{.01cm}\FeynmanBubble\hspace{.01cm}}
    =
    \alpha_1+\alpha_2
    +m_1^2\alpha_1^2
    +m_2^2\alpha_2^2
    +(m_1^2+m_2^2-p^2)\alpha_1\alpha_2\, ,
    \label{eq:bubble-graph-polynomial}
\end{equation}
and in generic kinematics the critical-point equations~\eqref{eq:critical-point-equations} admit three isolated solutions:
\begin{align}
    \sigma_1
    &=
    \left\{-\frac{1}{2m_1^2},\,0\right\}\, ,
    \label{eq:bubble-sigma1}
    \\
    \sigma_2
    &=
    \left\{0,\,-\frac{1}{2m_2^2}\right\}\, ,
    \label{eq:bubble-sigma2}
    \\
    \sigma_3
    &=
    \left\{
        \frac{p^2-m_1^2+m_2^2}{P},\,
        \frac{p^2+m_1^2-m_2^2}{P}
    \right\}\, ,
    \label{eq:bubble-sigma3}
\end{align}
where $P=\bigl(p^2-(m_1+m_2)^2\bigr)
     \bigl(p^2-(m_1-m_2)^2\bigr)$. Here, $\sigma_1$ lies in sector $\{1\}$, $\sigma_2$ lies in sector $\{2\}$, and $\sigma_3$ lies in sector $\{1,2\}$. Evaluating the graph polynomial at these critical points gives
\begin{align}
    \mathcal{G}_{\hspace{.01cm}\FeynmanBubble\hspace{.01cm}}(\sigma_1)
    &=
    -\frac{1}{4m_1^2}\, ,
    \label{eq:bubble-graph-at-sigma1}
    \\
    \mathcal{G}_{\hspace{.01cm}\FeynmanBubble\hspace{.01cm}}(\sigma_2)
    &=
    -\frac{1}{4m_2^2}\, ,
    \label{eq:bubble-graph-at-sigma2}
    \\
    \mathcal{G}_{\hspace{.01cm}\FeynmanBubble\hspace{.01cm}}(\sigma_3)
    &=
    \frac{p^2}
    {P}\, .
    \label{eq:bubble-graph-at-sigma3}
\end{align}

The critical-point-counting criterion can now be applied directly. On the locus $m_1=0$, the $\alpha_1$ coordinate of  $\sigma_1$ becomes infinite, so the critical point in the sector $\{1\}$ is lost. In a similar way, at $m_2=0$ the critical point in the sector $\{2\}$ is lost. At the threshold and pseudo-threshold loci $p^2=(m_1\pm m_2)^2$, $P=0$ and thus the critical point $\sigma_{3}$ is lost in sector $\{1,2\}$. Finally, when $p^2=0$, $\sigma_3$ is again lost because $\mathcal{G}_{\hspace{.01cm}\FeynmanBubble\hspace{.01cm}}(\sigma_3)=0$. These results are summarized in Table~\ref{tab:bubble_critical_points}, where we see that we have reproduced the sectors that were found in Table~\ref{tab:bubble-onshell-lines}, where the latter were found by explicitly solving the Landau equations. Once the corresponding on-shell sets $S_{\lambda,i}$ have been identified in this way, the hierarchical principle can be applied directly to determine which sequential discontinuities must vanish.

\begin{table}[t]
	\centering
	\begin{tabular}{c|c|c}
		\textbf{Singularity} & \textbf{Critical-Point Behavior} & \textbf{Sectors with Vanishing Cycles} \\
		\hline
		$m_1=0$ 
		& $\mathcal{G}_{\hspace{.01cm}\FeynmanBubble\hspace{.01cm}}(\sigma_1) \to \infty$ 
		& $\{1\}$ \\
		
		$m_2=0$ 
		& $\mathcal{G}_{\hspace{.01cm}\FeynmanBubble\hspace{.01cm}}(\sigma_2) \to \infty$  
		& $\{2\}$ \\
		
		$p^2=(m_1\pm m_2)^2$ 
		& $\mathcal{G}_{\hspace{.01cm}\FeynmanBubble\hspace{.01cm}}(\sigma_3) \to \infty$  
		& $\{1,2\}$ \\
		
		$p^2=0$ 
		& $\mathcal{G}_{\hspace{.01cm}\FeynmanBubble\hspace{.01cm}}(\sigma_3) \to 0$ 
		& $\{1,2\}$
	\end{tabular}
	\caption{For each singular hypersurface, at least one of the critical points must cease to exist due to the fact that the corresponding critical value vanishes or becomes infinite. By tracking which sector(s) lose critical points, we can determine which sector(s) the corresponding singularity arises in.}
	\label{tab:bubble_critical_points}
\end{table}

\subsection{Regulated Critical Point Counting} \label{sec:regulated_counting}

In degenerate examples, subtleties can arise that complicate the use of the critical point equations. For instance, non-isolated families of solutions can arise, such as curves. Clearly, counting the (infinite) number of solutions in these cases does not track the information we are after. Furthermore, the presence of these higher-dimensional varieties has been linked to the existence of so-called magic relations~\cite{Crisanti:2026rbc}, which break the traditional notion of sectors. This makes the implications of the hierarchical principle ambiguous, even if more refined methods are used to compute the true number of vanishing cycles.

Fortunately, these situations are straightforward to detect, and \DiscKosky{} always flags the presence of non-isolated solutions. When such solutions occur in generic kinematics, more work is needed to identify a basis of integration cycles, and we refrain from making genealogical predictions. However, when non-isolated solutions only arise in special kinematic limits, useful predictions can still sometimes be made, as we will discuss in Section~\ref{sec:non_repeating}.

A second, more subtle issue is that solutions to~\eqref{eq:critical-point-equations} can undercount the true number of critical points when additional solutions occur at infinity~\cite{Bitoun:2017nre,Boels:2015yna}. One way to detect such undercounting is to consider the \textit{regulated} critical point equations \cite{Frellesvig:2019uqt,Frellesvig:2020qot,Huh_2013}, which identify the set of critical points if one takes all of the $\nu_e$ to be generic complex numbers. In the Lee--Pomeransky representation, this choice gives rise to the twist factor
\begin{equation}
    u_{\vec{\nu}}(\vec{\alpha})
    =
    \mathcal{G}(\vec{\alpha},\vec{s})^{-D/2}
    \prod_{e=1}^{N}\alpha_e^{\nu_e-1}\, ,
    \label{eq:regulated-twist}
\end{equation}
which means the critical point equations become
\begin{equation}
    \partial_{\alpha_e}\log u_{\vec{\nu}}
    =
    \frac{\nu_e-1}{\alpha_e}
    -
    \frac{D}{2}
    \frac{\partial_{\alpha_e}\mathcal{G}}{\mathcal{G}}
    =0,
    \qquad e=1,\ldots,N\, .
\end{equation}
This gives us the so-called regulated critical point equations:
\begin{tcolorbox}[
    enhanced,
    arc=2mm,
    boxrule=0.6pt,
    left=2mm,
    right=2mm,
    top=1.2mm,
    bottom=1.2mm,
    colback=white,
    colframe=black,
]
    \textbf{Regulated Critical Point Equations:}
    \begin{equation}
    \begin{aligned}
        (\nu_e-1)\,\mathcal{G}
        -
        \frac{D}{2}\,
        \alpha_e\,\partial_{\alpha_e}\mathcal{G}
        &=0,
        \qquad e=1,\ldots,N. \\
        \alpha_1\cdots\alpha_N \mathcal{G} &\neq 0,\infty
        \label{eq:regulated-critical-point-equations}
    \end{aligned}
    \end{equation}
\end{tcolorbox}
These equations provide a more reliable count of the total number of independent cycles, albeit for a slightly more general class of Feynman integrals than one would normally encounter in practical calculations (due to the non-integer powers in the integrand). However, because individual Feynman parameters are not allowed to vanish (since the $\alpha_e =0$ hypersurfaces appear in the twist factor), we lose the natural stratification of critical points into sectors that we had via~\eqref{eq:critical_point_sector}. Nevertheless, the total number of regulated critical points can be compared with the number of unregulated critical points; when these two numbers match, the computational pipeline outlined in Section~\ref{sect:critical-point-counting} can be trusted~\cite{Chestnov:2026mpo,Huh_2013,Fevola:2023fzn,matsubaraheo2025hypergeometricdiscriminants}. It is important to note that this check need only be performed once, in generic kinematics, in order to guarantee that \DiscKosky{} makes no false predictions. Namely, since the unregulated critical point equations can only undercount the number of critical points (not overcount), using unregulated counting to probe special kinematic limits only risks identifying spurious drops in the number of critical points, not obscuring genuine drops. As such, there is no risk of ruling out sequences of discontinuities that are actually allowed.\footnote{More details and examples of these subtleties can be found in Appendix A of~\cite{Chestnov:2026mpo}.}

{\emergencystretch=2em
By default, \DiscKosky{} automatically implements both these checks---testing for higher-dimensional solutions and verifying that the regulated and unregulated critical-point counts agree in generic kinematics; it makes no genealogical predictions if either check fails (unless requested to do so, via options we will introduce in the next section). This ensures that any results produced by the program are conservative. Even so, as we will see in Section~\ref{sec:examples}, we are able to generate hundreds of genealogical predictions for many two- and three-loop Feynman integrals, often saturating the power of the hierarchical principle.\par}

\subsection{Efficiently Counting Critical Points}\label{sec:crit-point-theory}

As we have now seen, genealogical constraints can be derived as long as one is able to reliably count the number of solutions to the critical-point equations~\eqref{eq:critical-point-equations}. While this can be done by solving the critical point equations directly, such an approach involves more work than is necessary. A more efficient strategy is to recast the critical-point equations as an ideal. For a sector $S\subseteq\{1,\ldots,N\}$, we first define the restricted graph polynomial
\begin{equation}
    \mathcal{G}_S
    =
    \left.
    \mathcal{G}
    \right|_{\alpha_e=0,\;e\notin S}\, .
\end{equation}
The critical-point equations in this sector can then be encoded in the ideal
\begin{equation}
    \mathcal{J}_S
    =
    \left\langle
        \partial_{\alpha_e}\mathcal{G}_S\ \big|\ e\in S,
        \,
        1-\alpha_0\mathcal{G}_S
    \right\rangle ,
    \label{eq:crit-point-ideal}
\end{equation}
where $\alpha_0$ is an auxiliary variable introduced to enforce the condition $\mathcal{G}_S\neq0$. The number of critical points in a sector $S$ (counted with multiplicity) is then given by the \textit{degree} of $\mathcal{J}_S$,
\begin{equation}
	N_{S} = \deg(\mathcal{J}_S)\, .
\end{equation}
For a zero-dimensional ideal, $\deg(\mathcal{J}_S)$ is given by the dimension of the corresponding quotient ring, which can systematically be computed from the standard monomials of a Gr\"obner basis~\cite{Sturmfels2002,CoxLittleOShea2025}. While in general, the construction of Gr\"obner bases can be computationally expensive (particularly when the polynomial coefficients have complicated kinematic dependence), we are here only interested in the number of solutions at a generic kinematic point. We can therefore specialize the kinematic variables to generic values in a finite field, and perform the entire Gr\"obner-basis computation over that field.\footnote{Using Gr\"obner bases to find the full solution set of a system of equations also requires more expensive monomial orderings (such as lexicographic ordering) and back-substitution steps. Neither is needed for the computation of ideal degrees.} This avoids the growth of large rational functions in the kinematic variables that would otherwise occur during a symbolic computation by replacing the coefficient arithmetic by inexpensive finite-field operations.

In practice, \DiscKosky{} uses \texttt{msolve} as the underlying engine for computing Gr\"obner bases and ideal degrees. It sets each of the kinematic parameters to random integer numerical values, and uses a finite field of order $\sim 2^{31}$ (the largest set of primes that \texttt{msolve} can accept). This ensures that the possibility of unlucky primes is minimized, whilst maintaining the efficient computation strategy outlined above. Taken together, this renders the ideals that show up in two- and three-loop Feynman integrals at (and beyond) the current computational frontier extremely manageable~\cite{Chestnov:2025svg,Chestnov:2026mpo}. 

The computation of Gr\"obner bases over prime fields also serves a second crucial purpose, by allowing us to specialize to complicated kinematic hypersurfaces without introducing algebraic roots. The na\"ive approach to specialising to a given $\lambda(\vec{s})=0$ is to substitute random integer values for all but one kinematic variable, and then solve for the remaining variable. However, if this last variable appears quadratically (or worse), this final substitution will generally introduce radicals, which slow down the Gr\"obner basis computation significantly. 

In finite fields, this undesirable behavior can be avoided by exploiting the fact that polynomials often factor over large prime fields. Indeed, for a polynomial $\lambda(s)$ of degree $d$, the probability that it will contain a linear factor over a prime field $p$ is at least $\frac{1}{2}$ as long as $p\gg d$. Thus, by trying multiple candidate prime values, $\lambda(\vec{s})=0$ can be enforced without introducing radicals.\footnote{In practice, \DiscKosky{} can run through up to 400 different 31-bit primes, ensuring that the probability of not finding a factorization is vanishingly small.} As a quick example, consider the polynomial
\begin{equation}
	\lambda(s,t) = s^2 + 5st + t^2\,.
\end{equation}
Since $\lambda(s,t)$ is irreducible over the integers, solving for either $s$ or $t$ requires introducing radicals. However, a direct computation shows that
\begin{equation}
    \lambda(s,t)
    \equiv
    \bigl(s+516692786\,t\bigr)
    \bigl(s+1630790866\,t\bigr)
    \pmod{2^{31}-1}\, .
\end{equation}
As such, there exist integer combinations of $\{s,t\}$ that enforce $\lambda(s,t)\equiv0\pmod{2^{31}-1}$, for example $t = 1783515762$ and $s = 846172908$. These values can thus be substituted into $\mathcal{J}$ to ensure that $\lambda(s,t)=0$ throughout the Gr\"obner basis computation, and thus also in the evaluation of $\deg(\mathcal{J})$.
\subsection{Summary}\label{sect:theory-summary}
We can now summarize the main ingredients needed to derive genealogical constraints. For each candidate singularity $\lambda(\vec{s})=0$, we first determine the on-shell sets $S_{\lambda,i}$ using the critical-point-counting criterion. Namely, we count the isolated solutions to~\eqref{eq:critical-point-equations} sector by sector, both in generic kinematics and after restricting to $\lambda=0$; each sector in which the critical-point count decreases identifies a sector in which a singularity can arise as $\lambda \to 0$. Once these sets are known, the implications of the hierarchical principle can be easily deduced using~\eqref{eq:hierarchical-principle-background}.

The unregulated critical-point equations must, however, be treated with care. We therefore check for non-isolated solutions and compare the total number of critical points found via the unregulated critical point equations with the number obtained from regulated critical point equations. When these two counts agree in generic kinematics, we can reliably use the unregulated result to determine the sets $S_{\lambda,i}$.

In practice, \DiscKosky{} computes the number of critical points using Gr\"obner bases over finite fields. The kinematic variables are specialized to generic numerical values, and the critical-point count is compared between generic kinematics and the candidate singular surface. As we will illustrate in Section~\ref{sec:examples}, this strategy is extremely efficient, even for examples that go well beyond the current state of the art.

%% file: sections/quick_starter_guide.tex
\section{\DiscKosky{} Quick Start Guide}\label{sec:quickstart}
In this section we present the most important commands from the \DiscKosky{} package. More extensive documentation can be accessed once the package has been installed, via the ``Open Documentation'' button that appears when the package is loaded.
\subsection{Installation and Updates}
\DiscKosky{} is a paclet, and can be installed (or updated) directly from a Mathematica session with the command:
\begin{lstlisting}
ResourceFunction["GitHubInstall"]["giu989","DiscKosky"]
\end{lstlisting}
It can also be installed directly via the associated GitHub repository \githubrepo. Once installed, the package should be loadable as:
\begin{lstlisting}
<<DiscKosky`
\end{lstlisting}
The location of the installation can be found with
\begin{lstlisting}
PacletFind["DiscKosky"][[1,1]]["Location"];
\end{lstlisting}
It is highly recommended that \textsc{msolve} also be installed \cite{berthomieu} for the required Gr\"obner basis calculations. If installed in a standard path, \DiscKosky{} will detect and use \textsc{msolve} automatically. \textsc{msolve} is available through many standard package managers such as \textsc{apt} or \textsc{homebrew}.

\DiscKosky{} can be uninstalled by running
\begin{lstlisting}
PacletUninstall["DiscKosky"]
\end{lstlisting}
\subsection{Core Usage}
\label{sec:core_usage}

\begin{figure}
    \centering
           \begin{tikzpicture}[scale=1]
                \coordinate (a) at (-.8,.8);
                \coordinate (b) at (.8,.8);
                \coordinate (c) at (.8,-.8);
                \coordinate (d) at (-.8,-.8);
                \draw[] (a) to  node[midway, above] {$\alpha_1$}  (b);
                \draw[] (b) to  node[midway, right] {$\alpha_2$} (c);
                \draw[] (c) to  node[midway, below] {$\alpha_3$}  (d);
                \draw[] (d) to  node[midway, left] {$\alpha_4$}  (a);
                \draw[] (a) --++ (150:1);
                \draw[] (a) --++ (120:1);
                \draw[] (b) --++ (45:1);
                \draw[] (c) --++ (-30:1);
                \draw[] (c) --++ (-60:1);
                \draw[] (d) --++ (-135:1);
                
           \node[] at (-1.49,1.89) { $p_2$};
\node[] at (-1.95,1.25) { $p_1$};
\node[] at (-1.8,-1.7) { $p_6$};
\node[] at (1.7,-1.8) { $p_5$};
\node[] at (2,-1.3) {$p_4$};
\node[] at (1.9,1.8) {$p_3$};
            \end{tikzpicture}
    \caption{The two-mass easy box.}
    \label{fig:2m-easy}
\end{figure}
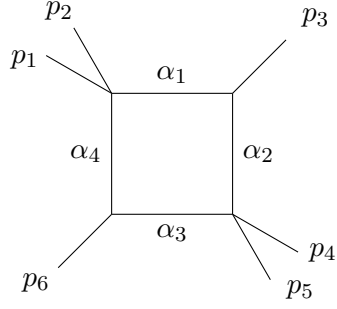

{\emergencystretch=2em
\DiscKosky{} provides three main high-level functions for predicting sequential discontinuities---\texttt{CheckDoubleDiscontinuities}, \texttt{CheckSequentialDiscontinuity} and \texttt{SectorDropGraph}. The three inputs needed to apply these functions to a given Feynman integral are the Lee--Pomeransky polynomial, a list of the integration variables, and a candidate set of singular loci in the space of external parameters. For example, if we consider the two-mass easy box shown in Figure~\ref{fig:2m-easy}, these ingredients are given by\par}
\begin{lstlisting}
g = x1+x2+x3-s23 x1 x3-m2^2 x2 x3+x4-m1^2 x1 x4-s12 x2 x4;
vars = {x1,x2,x3,x4};
sings = {s23,m1,m2,s12,-m1^2+s23,-m2^2+s23,m2^2-s12,m1^2-s12,m1^2 m2^2-s12 s23,
    -m1^2-m2^2+s12+s23};
\end{lstlisting}
Given these arguments, \texttt{CheckDoubleDiscontinuities} produces a matrix that tells the user which ordered sequences of discontinuities can appear in the Feynman integral. For instance, if we run
\begin{lstlisting}
dM1 = CheckDoubleDiscontinuities[g,vars,sings];
\end{lstlisting}
then the output matrix will be
\begin{equation}\label{eq:dM1}
    \texttt{dM1}=\left[
\begin{array}{cccccccccc}
 \True & \False & \False & \False & \texttt{Ind} & \texttt{Ind} & \False & \False & \True & \True \\
 \False & \True & \False & \False & \texttt{Ind} & \False & \False & \texttt{Ind} & \True & \True \\
 \False & \False & \True & \False & \False & \texttt{Ind} & \texttt{Ind} & \False & \True & \True \\
 \False & \False & \False & \True & \False & \False & \texttt{Ind} & \texttt{Ind} & \True & \True \\
 \False & \False & \False & \False & \texttt{Ind} & \False & \False & \False & \True & \True \\
 \False & \False & \False & \False & \False & \texttt{Ind} & \False & \False & \True & \True \\
 \False & \False & \False & \False & \False & \False & \texttt{Ind} & \False & \True & \True \\
 \False & \False & \False & \False & \False & \False & \False & \texttt{Ind} & \True & \True \\
 \False & \False & \False & \False & \False & \False & \False & \False & \True & \True \\
 \False & \False & \False & \False & \False & \False & \False & \False & \True & \True \\
\end{array}
\right]\,,
\end{equation}
where \texttt{Ind} is short for \texttt{Indeterminate} (and indicates that a higher-dimensional solution to the critical point equations has been found). The entry in the $i^\text{th}$ row and $j^\text{th}$ column of this matrix, namely \texttt{dM1[[i,j]]}, tells us whether a discontinuity with respect to the $j^\text{th}$ singularity can follow a discontinuity with respect to the $i^\text{th}$ singularity. Namely, if \texttt{dM1[[i,j]]} is \texttt{False}, this indicates that there exists a genealogical constraint of the form
\begin{equation}
\text{Disc}_{\texttt{sings[[j]]}}\,\, \cdots\,\, \text{Disc}_{\texttt{sings[[i]]}}\,\,\cdots\,\,\mathcal{I}_{\twomasseasybox} =0\, .
\end{equation}
Conversely, if \texttt{dM1[[i,j]]} is \texttt{True}, such sequences of discontinuities are not ruled out by our method. 

When \texttt{dM1[[i,j]]} is \texttt{Indeterminate}, there are two ways one can proceed. Most directly, these entries indicate that the algorithm described in Section~\ref{sect:theoretical-background} cannot be applied, so the conservative option is thus to treat such a result as equivalent to \texttt{True}. However, as we will show in Section~\ref{sec:non_repeating}, more can be said in many of the cases in which \texttt{Indeterminate} solutions arise (such as when a sector that has no critical points in generic kinematics becomes \texttt{Indeterminate} in a singular limit). These more refined predictions can be requested in \DiscKosky{} by using the option \texttt{"RefineIndeterminates"}, which can be passed to all of the core \DiscKosky{} functions:
\begin{lstlisting}
dM2 = CheckDoubleDiscontinuities[g,vars,sings,"RefineIndeterminates"->True];
\end{lstlisting}
\begin{equation}\label{eq:dM2}
    \texttt{dM2}=\left[
\begin{array}{cccccccccc}
 \True & \False & \False & \False & \True & \True & \False & \False & \True & \True \\
 \False & \True & \False & \False & \True & \False & \False & \True & \True & \True \\
 \False & \False & \True & \False & \False & \True & \True & \False & \True & \True \\
 \False & \False & \False & \True & \False & \False & \True & \True & \True & \True \\
 \False & \False & \False & \False & \False & \False & \False & \False & \True & \True \\
 \False & \False & \False & \False & \False & \False & \False & \False & \True & \True \\
 \False & \False & \False & \False & \False & \False & \False & \False & \True & \True \\
 \False & \False & \False & \False & \False & \False & \False & \False & \True & \True \\
 \False & \False & \False & \False & \False & \False & \False & \False & \True & \True \\
 \False & \False & \False & \False & \False & \False & \False & \False & \True & \True \\
\end{array}
\right]\,,
\end{equation}
{\emergencystretch=2em
Compared to \texttt{dM1}, four new singularities (\texttt{sings[[5]]}, \texttt{sings[[6]]}, \texttt{sings[[7]]}, and \texttt{sings[[8]]}) can now be identified as giving rise to non-repeating discontinuities, using the ideas presented in Section~\ref{sec:non_repeating}. Any \texttt{Indeterminate} entries which the algorithm from Section~\ref{sec:non_repeating} cannot refine are conservatively set to \texttt{True}.\par}

The function \texttt{CheckSequentialDiscontinuity} works in a very similar manner, except that it tests for a specific string of discontinuities of any length. It takes four inputs---the same three as above, and a specific sequence to test. For example,
\begin{lstlisting}
seq = {3,9,7}
CheckSequentialDiscontinuity[g, vars, sings, seq, "RefineIndeterminates" -> True]
(*Out: False*)
\end{lstlisting}
implies that
\begin{equation}
    \text{Disc}_{\texttt{sings[[7]]}}\,\, \cdots\,\, 
    \text{Disc}_{\texttt{sings[[9]]}}\,\, \cdots\,\,
    \text{Disc}_{\texttt{sings[[3]]}}\,\,\cdots\,\,\mathcal{I}_{\twomasseasybox} = 0 \,.
\end{equation}
The information given by the above functions can also be displayed graphically with the command \texttt{SectorDropGraph}; running
\begin{lstlisting}
graph = SectorDropGraph[g, vars, sings, "RefineIndeterminates" -> True];
\end{lstlisting}
one gets
\begin{equation} \label{eq:disckosky_graph}
    \text{\texttt{graph}} =
    \vcenter{\hbox{\includegraphics[width=0.8\linewidth]{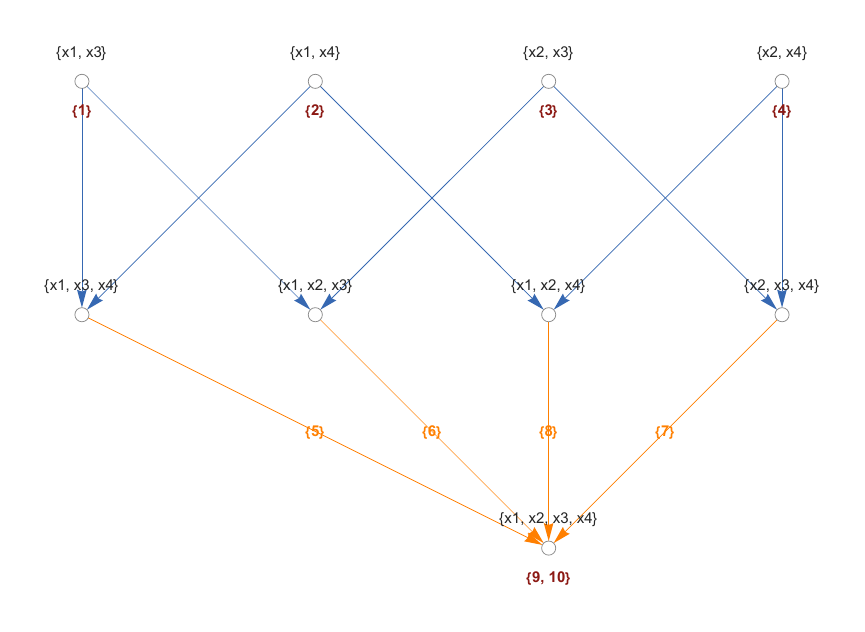}}}\,.
\end{equation}
Each node in this graph represents the sector of the Feynman integral corresponding to the integration variables that appear above it (for instance, the top left node, labeled by $\{x_1, x_3\}$, corresponds to the sector in  which $x_2=x_4=0$); the (indices of the) singularities that appear in each of these sectors are displayed below the node. Allowed sequences of discontinuities can be read off of this graph by either choosing a singularity at the current node, or one that can be reached by following any sequence of directed arrows. If the option \texttt{"RefineIndeterminates"} is enabled, some singularities may be placed on edges (marked in orange). These represent non-repeating discontinuities. 
\subsection{Other Useful Functions}
\DiscKosky{} also exposes to the user some lower-level functions that can be used to analyze possible sequences of discontinuities. The first of these is \texttt{CountSectorsUnregulated}, which computes the number of solutions to the critical point equations~\eqref{eq:critical-point-equations} in each sector. \texttt{CountSectorsUnregulated} takes as input a Lee--Pomeransky polynomial, a list of integration variables, and a set of propagators to cut. The result is the number of critical points in each sector above the specified cut.\footnote{The output of this function mirrors closely that of \textsc{Mint}~\cite{Lee_2013}.} For example, sticking with the two-mass easy box example,
\begin{lstlisting}
cut = {};
data = CountSectorsUnregulated[g, vars, cut];
data[[1]] (*Out: 5*)
data[[2]] (*Out: {0, 0, 0, 0, 0, 0, 1, 1, 1, 1, 0, 0, 0, 0, 0, 1}*)
data[[3]] (*Out: {{}, {x1}, {x2}, {x3}, {x4}, {x1, x2}, {x1, x3}, {x1, x4}, {x2, 
  x3}, {x2, x4}, {x3, x4}, {x1, x2, x3}, {x1, x2, x4}, {x1, x3, 
  x4}, {x2, x3, x4}, {x1, x2, x3, x4}}*)
\end{lstlisting}
The first output corresponds to the total number of critical points (summed over all sectors that are consistent with the specified cut), while the second indicates how these critical points are distributed through the different sectors. The third output provides the list of sectors that label the critical point counts from the second output. 

If one specifies a non-empty \texttt{cut}, the function will simply ignore any sectors that do not survive the support of that cut. For example:
\begin{lstlisting}
cut = {x1, x2, x3, x4};
CountSectorsUnregulated[g, vars, cut] (*Out: {1, {1}, {{x1, x2, x3, x4}}}*)
\end{lstlisting}
In many cases, the number of these critical points can be interpreted as the number of master integrals in a given sector \cite{Lee_2013}. However, as discussed in Section~\ref{sect:theoretical-background}, this number is known to sometimes undercount \cite{Boels:2015yna,Bitoun:2017nre}. When combined with the possibility of magic relations \cite{Crisanti:2026rbc}, a naive interpretation of this result must be taken with caution. Explicit examples illustrating these problematic edge cases are presented in the package's documentation pages.

\texttt{CountSectorsUnregulated} also supports the computation of critical points on the support of kinematic constraints, via the \texttt{"Constraint"} option. For example,
\begin{lstlisting}
cut = {x1, x2, x3, x4};
data = CountSectorsUnregulated[g, vars, cut, "Constraint" -> sings[[10]]]
(*Out : {0, {0}, {{x1, x2, x3, x4}}}*)
\end{lstlisting}
shows that one critical point is lost in the top sector when the condition $\texttt{sings[[10]]} = s_{12} + s_{23}-m_1^2-m_2^2= 0$ is enforced on the external kinematics. This indicates that the corresponding Lefschetz thimble can play the role of a vanishing cycle when a discontinuity is computed with respect to the singularity at $s_{12} + s_{23}-m_1^2-m_2^2= 0$.

It is also possible to access directly which on-shell lines/sectors correspond to a singularity with the command \texttt{FindSingularitySectors}. This command takes as input a Lee--Pomeransky polynomial, a list of integration variables, the set of candidate Landau singularities as well as the index of the specific singularity to analyze. For example,
\begin{lstlisting}
FindSingularitySectors[g,vars,sings,10]
(*Out: {{x1, x2, x3, x4}}*)
\end{lstlisting}
returns that the singularity $s_{12} + s_{23}-m_1^2-m_2^2= 0$ is associated with (only) the top sector of the one-loop box.

{\emergencystretch=5em
Finally, \DiscKosky{} provides a public function for counting with analytic regulators \cite{Frellesvig:2019uqt}. More precisely, this function computes the absolute value of the Euler characteristic \mbox{$\lvert\chi(\mathbb{C}^n \setminus V(x_1 \cdots x_n \mathcal{G}))\rvert$}, which can always be interpreted as the number of master integrals of the (regulated) Feynman integral. In most cases, the output of \texttt{CountSectorsUnregulated} will agree with this more rigorous computation:\par}
\begin{lstlisting}
cut = {};
CountSectorsUnregulated[g, vars, cut][[1]] (*Out: 5*)
CountSectorsRegulated[g, vars, cut] (*Out: 5*)
\end{lstlisting}
{\raggedright
\texttt{CountSectorsRegulated} and \texttt{CountSectorsUnregulated} also support the \texttt{"Constraint"} option.\par}

%% file: sections/Examples.tex
\section{\emph{Exempla Admirabilia}}\label{sec:examples}

Having introduced the main functionality of \DiscKosky{}, we now illustrate its use in several examples. For the polylogarithmic integrals considered below, it is useful to phrase these constraints in terms of their symbol~\cite{Goncharov:2010jf}, which records the sequences of logarithmic integrations these integrals encode. Namely, constraints on the sequential discontinuities of such integrals translate into restrictions on the order in which different symbol letters (or $d\log$ arguments) can appear. For more details on the symbol of multiple polylogarithms, see for instance~\cite{Duhr:2014woa,Weinzierl:2022eaz}. Note, however, that genealogical constraints can just as easily be derived for Feynman integrals that evaluate to more complicated classes of special functions, such as those involving integrals over elliptic curves or Calabi--Yau manifolds~\cite{Bourjaily:2022bwx}.

We first consider two-loop families of Feynman integrals for which canonical differential equations are known, and show that in each case \DiscKosky{} predicts all possible hierarchical constraints between the rational letters in their symbol alphabets (leaving consideration of algebraic letters until Section~\ref{sec:algebraic-letters}). We then apply the same analysis to a three-loop example---the massive envelope diagram---for which no analytic result is currently known, and derive hundreds of new genealogical constraints. For each family of integrals, deriving these predictions takes approximately a minute (or less). Finally, we illustrate how \DiscKosky{} can be used to derive genealogical constraints on more general Euler integrals by applying it to examples that appear in the evaluation of cosmological correlators. Together, these examples demonstrate both the computational efficiency and practical reach of \DiscKosky{}.

\subsection{Two- and Three-Loop Feynman Integrals}

We begin by testing \DiscKosky{} on the one-mass hexabox families computed in~\cite{Abreu:2021smk}, and the two-mass pentabox families computed in~\cite{Abreu_2024}. For each configuration of external masses, we supply \DiscKosky{} with the corresponding Lee--Pomeransky polynomial and the set of logarithmic branch points that appear in each family.\footnote{More precisely, we include only logarithmic branch points that do not also arise as algebraic branch points. Genealogical constraints involving algebraic branch points can also be derived, but we defer their discussion to Section~\ref{sec:algebraic-letters}.} \DiscKosky{} then identifies the sectors in which each of these logarithmic singularities arises, and derives the corresponding set of hierarchical constraints. The results are summarized in Table~\ref{tab:some 2 loop rational predictions}.

\renewcommand{\arraystretch}{1.3}
\begin{table}[ht]
\centering
\begin{adjustbox}{max width=\textwidth}
\centering
\begin{tabular}{|c|c|c|c|c|}
\hline
\makecell{\textbf{Family of} \\ \textbf{Diagrams}} 
& \makecell{\textbf{Mass} \\ \textbf{Configuration}} 
& \makecell{\textbf{Pairs of}\\ \textbf{ Branch Points}} 
& \makecell{\textbf{Genealogical} \\ \textbf{Constraints}} 
& \makecell{\textbf{Timing} \\ \textbf{(s)}} \\
\hline
\multirow{3}{*}{
\adjustbox{width=2.5cm,valign=c}{
\centering\begin{tikzpicture}[scale=1]
    \coordinate (a) at (-.8,0);
    \coordinate (b) at (0,.8);
    \coordinate (c) at (.8,0);
    \coordinate (d) at (0,-.8);
    \coordinate (e) at (2.2, 1.2);
    \coordinate (f) at (3, 0);
    \coordinate (g) at (2.2, -1.2);

    \coordinate (a1) at (-1.6, 0);
    \coordinate (c1) at (1.6, 0);
    \coordinate (e1) at (2.6, 1.6);
    \coordinate (f1) at (3.6, 0);
    \coordinate (g1) at (2.6, -1.6);

    \draw(a) -- (b) -- (c) -- (d) -- cycle;
    \draw (b) -- (e) -- (f) -- (g) -- (d);
    \draw[color=Maroon, line width = 3] (a1) -- (a);
    \draw(c1) -- (c);
    \draw(e1) -- (e);
    \draw (f1) -- (f);
    \draw (g1) -- (g);

\end{tikzpicture}
}}

& zzz & $33^2=1089$ & 322 & 59.7 \\
\cline{2-5}
& mzz & $26^2=676$ &215 & 38.7 \\
\cline{2-5}
& zmz & $29^2=841$ & 188 & 49.0 \\
\hline
\multirow{6}{*}{
\adjustbox{width=3.5cm,valign=c}{
\begin{tikzpicture}[scale=0.8]
    \coordinate (a) at (-.8,.8);
    \coordinate (b) at (.8,.8);
    \coordinate (c) at (.8,-.8);
    \coordinate (d) at (-.8,-.8);
    \coordinate (e) at (2.2, 1.2);
    \coordinate (f) at (3, 0);
    \coordinate (g) at (2.2, -1.2);

    \coordinate (a1) at (-1.6, 1.6);
    \coordinate (d1) at (-1.6, -1.6);
    \coordinate (e1) at (3, 2);
    \coordinate (f1) at (4, 0);
    \coordinate (g1) at (3, -2);

    \draw (a) -- (b) -- (c) -- (d) -- cycle;
    \draw (b) -- (e) -- (f) -- (g) -- (c);
    \draw[color=RoyalBlue, line width = 2] (a1) -- (a);
    \draw[color=Maroon, line width = 2] (d1) -- (d);
    \draw (e1) -- (e);
    \draw (f1) -- (f);
    \draw (g1) -- (g);
\end{tikzpicture}
}}

& zzz & $34^2=1156$ &376 & 62.9 \\
\cline{2-5}
& zzm & $32^2=1024$ &343 & 57.8 \\
\cline{2-5}
& zmz & $35^2=1225$ &391 & 68.1 \\
\cline{2-5}
& mzz & $36^2=1296$ &544 & 67.0 \\
\cline{2-5}
& mzm & $33^2=1089$ &489 & 59.6 \\
\cline{2-5}
& mmz & $31^2=961$ &254 & 55.0 \\
\hline
\centering\begin{tikzpicture}[scale=0.8, baseline={([yshift=-0.5ex]current bounding box.center)}]
    \coordinate (a) at (-.8,.8);
    \coordinate (b) at (.8,.8);
    \coordinate (c) at (.8,-.8);
    \coordinate (d) at (-.8,-.8);

    \coordinate (a1) at (-1.3, 1.3);
    \coordinate (d1) at (-1.3, -1.3);
    \coordinate (b1) at (1.3,1.3);
    \coordinate (c1) at (1.3,-1.3);

    \draw[color=RoyalBlue, line width = 1.5] (a) -- (b) -- (c) -- (d) -- cycle;
    \draw[color=RoyalBlue, line width = 1.5] (a) -- (-0.1,0.1);
    \draw[color=RoyalBlue, line width = 1.5] (0.1,-0.1) -- (c);
    \draw[color=RoyalBlue, line width = 1.5] (b) -- (d);
    \draw[Maroon, line width = 1.5] (a1) -- (a);
    \draw[Maroon, line width = 1.5] (d1) -- (d);
    \draw[Maroon, line width = 1.5] (b1) -- (b);
    \draw[Maroon, line width = 1.5] (c1) -- (c);
\path ([yshift=0.1cm]current bounding box.north) ([yshift=-0.1cm]current bounding box.south);
\end{tikzpicture}
& -- & $31^2=961$ &342 & 21.8 \\
\hline
\end{tabular}
\end{adjustbox}
\caption{The genealogical constraints generated by \DiscKosky{} between the logarithmic branch points that appear in different mass configurations of the five-point one-mass hexabox and two-mass pentabox families. For each mass configuration, we report the total number of ordered pairs of rational letters (where we exclude rational letters that also appear as the argument of square roots), and the number of these pairs that are ruled out by \DiscKosky{}; the family labels follow the conventions of~\cite{Abreu:2021smk,Abreu_2024}. It has been checked that these genealogical constraints are obeyed to all orders in dimensional regularization, and that they exhaust the hierarchical restrictions that hold between rational letters in these families of integrals. We also report the number of genealogical constraints that \DiscKosky{} generates for the three-loop fully massive non-planar envelope diagram, using as input the set of singularities predicted by the function \texttt{SPQRLandau[]}~\cite{Chestnov:2026mpo}; these predictions cannot be checked because the differential equation for this integral is not yet known.}
\label{tab:some 2 loop rational predictions}
\end{table}

Across all of the one-mass hexabox and two-mass pentabox mass configurations, \DiscKosky{} prohibits approximately $33.4\%$ of the ordered pairs of rational letters from appearing. Despite the size of the alphabets involved, \DiscKosky{} is able to derive these constraints in approximately one minute per mass configuration. Note that these predictions include the identification of non-repeating letters (which will be discussed in more detail in the next section); for instance, the letters
\begin{equation}
    s_{15}-s_{34}
    \qquad\text{and}\qquad
    p_1^2-s_{12}-s_{15}+s_{23}+s_{34}\, 
\end{equation}
can appear at most once in any symbol term that appears in the mzz hexabox family.\footnote{In some cases, additional non-repeating discontinuity constraints can arise in the subsectors of an integral family. For example, the letter $s_{23}-s_{15}$ can appear at most once in each term of the symbol of the one-mass five-point double box considered in Section~\ref{sect:hexabox}, but this constraint does not hold for any of the full integral families listed in Table~\ref{tab:some 2 loop rational predictions}.} We have checked all of \DiscKosky{}'s predictions against the differential equation matrices that have been derived for each of these integral families~\cite{Abreu:2021smk,Abreu_2024}. In every case, the predictions made by \DiscKosky{} are obeyed to all orders in dimensional regularization; moreover, they exhaust the set of hierarchical restrictions that can be observed to hold between the rational symbol letters in these integrals. 

As an instructive exercise, we have also checked that the full set of singularities that appear in these integral families can be generated independently, using the function \texttt{SPQRLandau[]} presented in~\cite{Chestnov:2026mpo}. For each mass configuration, the singular loci obtained in this way agree exactly with the set of singularities that appear in the final answer (although this analysis does not yet help us distinguish between logarithmic and algebraic branch points).

Having validated \DiscKosky{} against these two-loop examples, we next apply it to the three-loop, fully massive non-planar envelope diagram, with equal external masses and equal internal masses.\footnote{This integral is of particular interest in the $S$-matrix bootstrap, where it provides the leading contribution to the multiparticle double spectral density and thus serves as a simple probe of multiparticle unitarity~\cite{Correia:2021etg,Tourkine:2023xtu}.} This integral is expected to have 31 distinct singular loci~\cite{Chestnov:2026mpo}, although no canonical differential equation or analytic result is currently available. Of the $31^2=961$ ordered pairs of discontinuities that might appear, \DiscKosky{} predicts that 342 are hierarchically forbidden. This computation takes just \(21.8\) seconds.

\subsection{Cosmological Correlators}\label{sec:correlators}

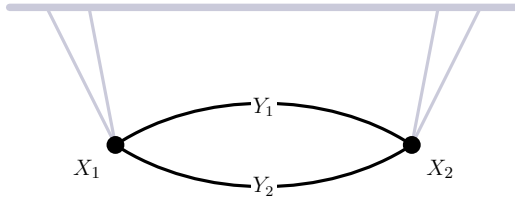
\begin{figure*}[t]
    \centering
    \scalebox{0.7}{\input{Figures/bubble_correlator.tex}}
    \caption{The bubble cosmological correlator diagram, which depends on the energies $X_i$ and $Y_e$ that flow into vertex $i$ and through edge $e$. The gray lines denote external particles that are detected on a future spatial hypersurface.}
    \label{fig:bubble_correlator}
\end{figure*}

\DiscKosky{} can also be used to study Euler integrals that involve multiple denominator factors, such as those that arise in the computation of cosmological correlators. As a first example, consider the energy integrals associated with the bubble diagram shown in Figure~\ref{fig:bubble_correlator}~\cite{Arkani-Hamed:2023kig,Arkani-Hamed:2023bsv,De:2023xue}:
\begin{equation}
\begin{aligned}
    {\cal I}_{\bubble}(X_i,Y_i)
    =
    \int_{x_i \ge 0}
    {\rm d}x_1\,{\rm d}x_2\,
    x_1^{T_1}x_2^{T_2}
    \left(
    \frac{1}{D_1D_2D_3D_5}
    +
    \frac{1}{D_1D_2D_4D_5}
    \right)\, ,
\end{aligned}
\label{eq:bubble_correlator}
\end{equation}
where
\begin{equation}
\begin{gathered}
    D_1 = x_1+X_1+Y_1+Y_2\, ,
    \qquad
    D_2 = x_2+X_2+Y_1+Y_2\, ,\\
    D_3 = x_1+x_2+X_1+X_2+2Y_1\, ,
    \qquad
    D_4 = x_1+x_2+X_1+X_2+2Y_2\, ,\\
    D_5 = x_1+x_2+X_1+X_2\, .
\end{gathered}
\end{equation}
Here, the external parameters $X_i$ and $Y_e$ represent the energies flowing into vertex $i$ and through edge $e$, respectively, while the exponents $T_i$ can take non-integer values.\footnote{In the cosmological correlator literature, the powers $T_i$ would usually be denoted by $\alpha_i$; we have not followed this convention here, to avoid confusing them with Feynman parameters.} For this reason, we treat $x_1^{T_1}x_2^{T_2}$ as a twist factor (in line with the discussion in Section~\ref{sec:regulated_counting}).

Unlike for flat-space Feynman integrals, where the loop momenta can be integrated out after introducing Feynman parameters, there is no general procedure for performing the \(x_i\) integrals in~\eqref{eq:bubble_correlator} after combining all $X_i$ and $Y_e$ dependence into a single denominator factor. We therefore work directly with products of the original $D_i$ denominator factors. Moreover,  due to the twist $x_1^{T_1}x_2^{T_2}$, none of the critical points of this integral can lie on the \(x_i=0\) integration boundaries; as such, the critical points do not naturally separate into different boundary strata. Instead, for any chosen collection of denominator factors, all critical points appear in the top sector, absent any distinguishing feature that allows us to determine which subset of denominators is actually responsible for each one. To recover this information, we consider each subset of denominator factors separately, and compare its critical-point count with those that arise when only a subset of the same denominator factors is present~\cite{Frellesvig:2019uqt,Frellesvig:2020qot,Brunello:2023rpq}. By subtracting off the number of critical points that already appear in subsectors, we can identify how many critical points arise for the first time in the current sector.

\begin{figure}\centering
\resizebox{\linewidth}{!}{\input{Figures/bubble_critical_point_figure}}
  \caption{The total number of critical points $C$ in the sector corresponding to each polynomial $\mathcal{D}_{\vec{\beta}}$, as well as the number of new critical points $\Delta C$ that the sector has not inherited from subsectors. Each of the sectors in which a new critical point appears is colored, and the arrows indicate which sectors inherit these critical points.}
  \label{fig:bubble_correlator_sectors_CPs}
\end{figure}
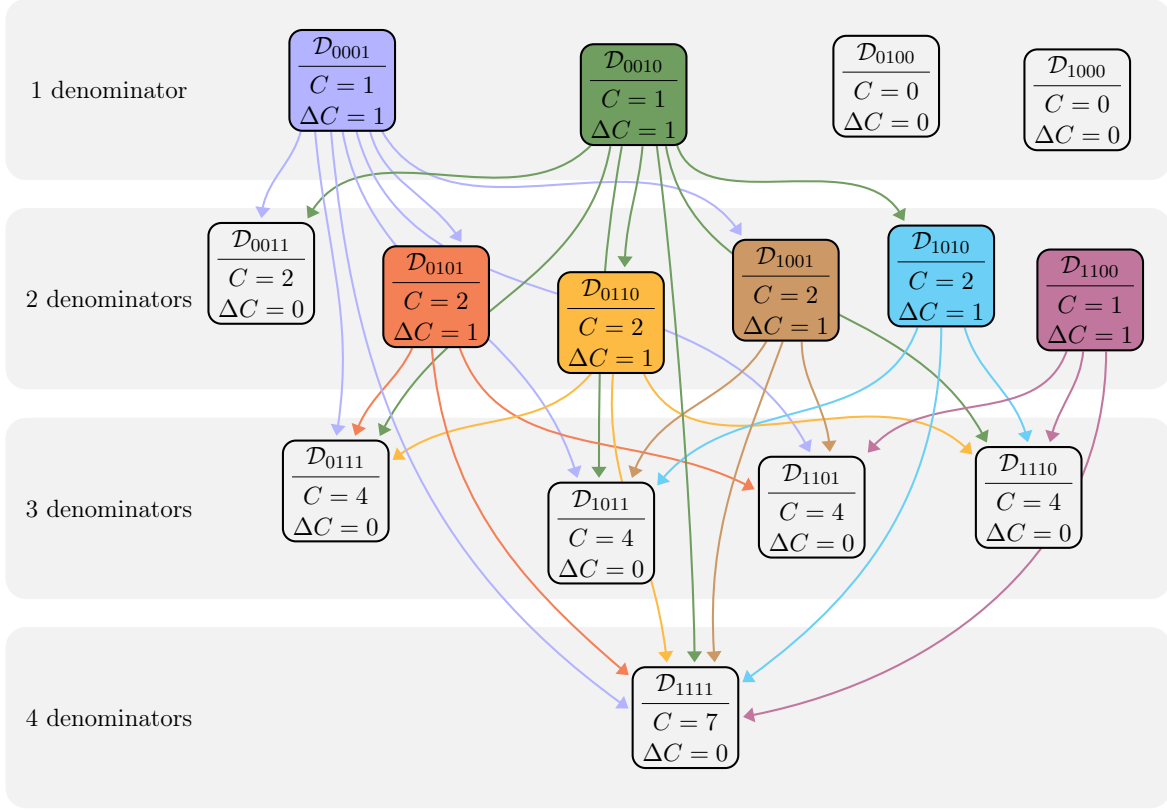

Let us demonstrate how this works in practice, focusing on just the first term that appears in the integrand of~\eqref{eq:bubble_correlator}. We consider the polynomial
\begin{equation}
\begin{aligned}
    {\cal D}_{\vec{\beta}}&=x_1^{-T_1} x_2^{-T_2}{\color{RoyalBlue}\underbrace{{\color{black}(x_1+X_1+Y_1+Y_2)}}_{\color{black} D_1}}^{\beta_1}{\color{RoyalBlue}\underbrace{{\color{black}(x_2+X_2+Y_1+Y_2)}}_{\color{black} D_2}}^{\beta_2} \\
    &\times {\color{RoyalBlue}\underbrace{{\color{black}(x_1+x_2+X_1+X_2+2Y_1)}}_{\color{black} D_3}}^{\beta_3}{\color{RoyalBlue}\underbrace{{\color{black}(x_1+x_2+X_1+X_2)}}_{\color{black} D_5}}^{\beta_4},
\end{aligned}
\end{equation}
where the exponents $\smash{\vec{\beta} = \{\beta_1,\beta_2,\beta_3,\beta_4\}}$ each take the value 0 or 1, and thus control which denominator factors are present. Evaluating the critical points across all possible $\smash{\vec{\beta}}$ sectors, one finds the critical point counts exhibited in Figure~\ref{fig:bubble_correlator_sectors_CPs}.\footnote{In practice, we substitute random integer values for $T_1$ and $T_2$ when plugging $\mathcal{D}_{\vec{\beta}}$ into \DiscKosky{}.} As seen there, we associate two quantities with each sector---the overall number of critical points $C$ as reported by \DiscKosky{}, and the number of new critical points $\Delta C$ that the sector has not inherited from its subsectors. More specifically, to determine $\Delta C$ in a given sector, we subtract from $C$ the values of $\Delta C$ that appear in all subsectors. As an example, consider sector $\vec{\beta} = \{1,1,1,0\}$, which contains a total of 4 critical points. This sector has a total of six subsectors, two of which have $\Delta C = 0$ ($\{1,0,0,0\}$ and $\{0,1,0,0\}$), and four of which have $\Delta C = 1$ ($\{0,0,1,0\}$, $\{1,1,0,0\}$, $\{1,0,1,0\}$, and $\{0,1,1,0\}$). Thus, all of the critical points that appear in this sector are already visible in subsectors, so we are able to conclude that no new critical points appear in this sector, which is to say that $\Delta C =0$ for $\{1,1,1,0\}$. Note that this is exactly what one should expect in a sector involving three denominator factors, given that the integral~\eqref{eq:bubble_correlator} involves only two integration variables. Indeed, the same analysis shows that $\Delta C=0$ for every three- and four-denominator sector in Figure~\ref{fig:bubble_correlator_sectors_CPs}; the critical points that \DiscKosky{} finds in these sectors are all inherited from lower sectors.

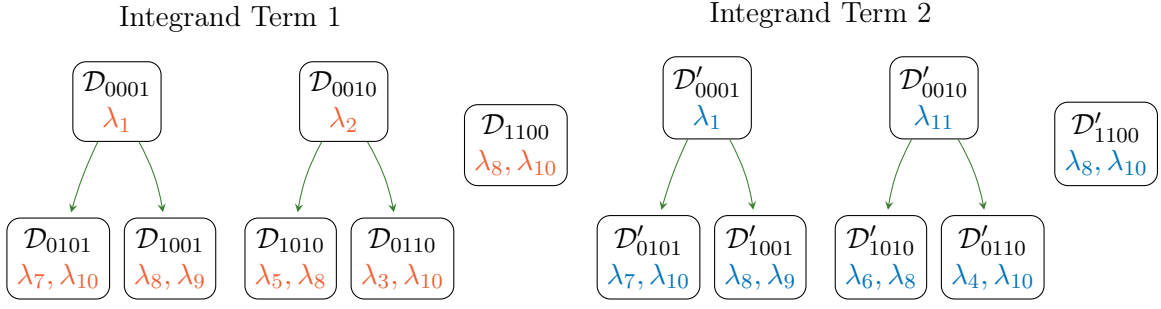
\begin{figure}
\centering
\input{Figures/cosmological_correlator_predictions_term_1}
\hspace{0cm}
\input{Figures/cosmological_correlator_predictions_term_2}
  \caption{The singularities that lead to a nonzero drop in $\Delta C$ in each sector for the bubble integral~\eqref{eq:bubble_correlator}. The arrows indicate which sequences of discontinuities are allowed by the hierarchical principle, as determined by the set of denominator factors that are involved in giving rise to each singularity---namely, after computing a discontinuity with respect to one of the singularities in a given sector, one can only compute a discontinuity with respect to a singularity in the same sector or a sector that can be reached by following these arrows. Note that there is no way to compute sequential discontinuities that require moving between sectors associated with different integrand terms. Any sequence of discontinuities that can appear in either term can appear in the bubble correlator integral.}
  \label{fig:bubble-correlator-hiearchical}
\end{figure}

To determine the sectors we should associate with a given kinematic singularity, we proceed as before and look for changes in the number of critical points sector by sector. Namely, we re-count $\Delta C$ in each sector after imposing the relevant kinematic constraint, and identify the sectors in which this number has decreased. This procedure mirrors closely that of \cite{Brunello:2023rpq} for computing Feynman integrals in each sector. In the case of the bubble integral~\eqref{eq:bubble_correlator}, there are 11 branch points we can check, corresponding to the following polynomials:
\begin{equation}\label{Correlators:discs}
\begin{aligned}
    \lambda_1 &= X_1 + X_2, & \lambda_2 &= X_1 + X_2 + 2 Y_1, & \lambda_3 &= -X_1 - Y_1 + Y_2, \\
    \lambda_4 &= X_1 - Y_1 + Y_2, & \lambda_5 &= -X_2 - Y_1 + Y_2, & \lambda_6 &= X_2 - Y_1 + Y_2, \\
    \lambda_7 &= -X_1 + Y_1 + Y_2, & \lambda_8 &= X_1 + Y_1 + Y_2, & \lambda_9 &= -X_2 + Y_1 + Y_2, \\
    \lambda_{10} &= X_2 + Y_1 + Y_2, & \lambda_{11} &= X_1 + X_2 + 2 Y_2. &
\end{aligned} 
\end{equation}
Since we are ultimately interested in which sequences of discontinuities can appear in the full integral, we also carry out the relevant analysis for the second term in~\eqref{eq:bubble_correlator}, which involves a different set of denominator factors, as encoded by the polynomial
\begin{equation}
\begin{aligned}
    {\cal D}^\prime_{\vec{\beta}}
    &=
    x_1^{-T_1}x_2^{-T_2}
    {\color{RoyalBlue}
    \underbrace{{\color{black}(x_1+X_1+Y_1+Y_2)}}_{\color{black}D_1}}^{\beta_1}
    {\color{RoyalBlue}
    \underbrace{{\color{black}(x_2+X_2+Y_1+Y_2)}}_{\color{black}D_2}}^{\beta_2}
    \\
    &\times
    {\color{RoyalBlue}
    \underbrace{{\color{black}(x_1+x_2+X_1+X_2+2Y_2)}}_{\color{black}D_4}}^{\beta_3}
    {\color{RoyalBlue}
    \underbrace{{\color{black}(x_1+x_2+X_1+X_2)}}_{\color{black}D_5}}^{\beta_4}.
\end{aligned}
\end{equation}
Tracking the sectors in which $\Delta C$ drops on the support of each singularity, we find the sector assignments summarized in Figure~\ref{fig:bubble-correlator-hiearchical}. There, we see that the correspondence between singularities and denominator sectors is not one-to-one. For instance, while some singularities such as $\lambda_4$ appear in just a single sector, others such as $\lambda_8$ appear in multiple sectors. This distinction becomes important when considering longer sequences of discontinuities, since a singularity at $\lambda=0$ that arises in multiple sectors can give rise to nonzero discontinuities that involve localizing the integration contour to different sets of denominator factors. 

For instance, in the first term, we see that a discontinuity with respect to $\lambda_1=0$ can be followed by a discontinuity with respect to $\lambda_8=0$ by moving between the sectors associated with $\mathcal{D}_{0001}$ and $\mathcal{D}_{1001}$. Similarly, a discontinuity with respect to $\lambda_8=0$ can be followed by a discontinuity with respect to $\lambda_{10}=0$ due to the sector associated with $\mathcal{D}_{1100}$. Thus, in general, we can have
\begin{align}
    \text{Disc}_{\lambda_8} \cdots \text{Disc}_{\lambda_1} \cdots {\cal I}_\bubble&\neq 0\, , \label{eq:bubble_allow_discs_1} \\
    \text{Disc}_{\lambda_{10}} \cdots \text{Disc}_{\lambda_8} \cdots {\cal I}_\bubble&\neq 0\, . \label{eq:bubble_allow_discs_2}
\end{align}
However, because the discontinuity with respect to $\lambda_8=0$ that appears in~\eqref{eq:bubble_allow_discs_1} requires localizing the integration contour to $D_1=0$ and $D_5=0$, while the discontinuity with respect to $\lambda_8=0$ in~\eqref{eq:bubble_allow_discs_2} requires localizing the integration contour to $D_1=0$ and $D_2=0$, these sequences of discontinuities are genealogically incompatible. That is, 
\begin{align}
    \text{Disc}_{\lambda_{10}} \cdots \text{Disc}_{\lambda_8} \cdots \text{Disc}_{\lambda_1} \cdots {\cal I}_\bubble&= 0\, . 
\end{align}
Note that the same phenomenon also occurs for flat-space Feynman integrals; it just becomes especially apparent here since we have more denominator factors than integration variables.

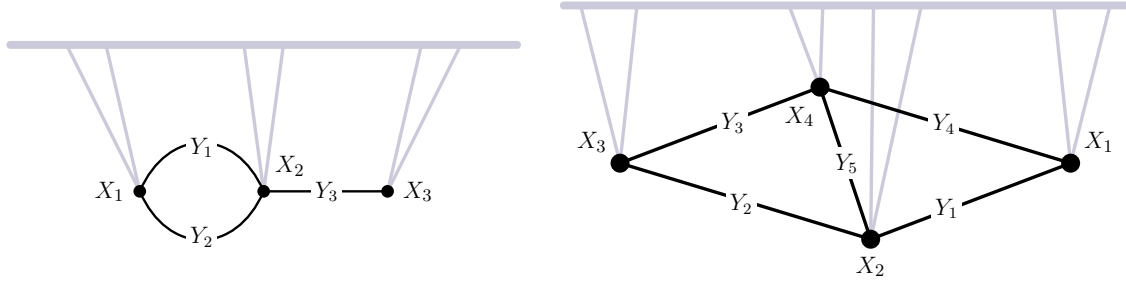
\begin{figure}[t]
    \centering
    \resizebox{.99\textwidth}{!}{%
        \input{Figures/ball_chain_correlator.tex}%
        \input{Figures/acnode_correlator.tex}%
    }

\caption{The chain-bubble correlator diagram (left) and acnode correlator diagram (right). Gray lines denote external particles that are detected on a future spatial hypersurface, while black lines denote internal propagators.}\label{fig:additional_correlator_examples}
\end{figure}

\begin{table}[b]
\centering

\definecolor{propagatorgray}{RGB}{202,202,218}

\begin{tabular}{|c||c|c|c|}
\hline
\makecell{\textbf{Cosmological Correlator}\\ \textbf{Diagram}}
&
\makecell{\textbf{Number of}\\ \textbf{Singular Loci}}
&
\makecell{\textbf{Ordered Pairs}\\ \textbf{of Singularities}}
&
\makecell{\textbf{Genealogical}\\ \textbf{Constraints}}
\\
\hline


\raisebox{-0.5\height}{
\resizebox{!}{1.45cm}{
\begin{tikzpicture}[
    line cap=round,
    line join=round
]

\path[use as bounding box] (-5.2,-1.8) rectangle (5.2,3.0);

\coordinate (X1) at (-2.8,0);
\coordinate (X2) at ( 2.8,0);

\def\futureheight{2.6}

\draw[
    propagatorgray,
    line width=4pt
]
(-4.8,\futureheight) -- (4.8,\futureheight);

\draw[propagatorgray, line width=1.7pt]
    (X1) -- (-4.1,\futureheight);

\draw[propagatorgray, line width=1.7pt]
    (X1) -- (-3.3,\futureheight);

\draw[propagatorgray, line width=1.7pt]
    (X2) -- (3.3,\futureheight);

\draw[propagatorgray, line width=1.7pt]
    (X2) -- (4.1,\futureheight);

\draw[black, line width=1.7pt]
(X1)
.. controls (-1.2,1.05) and (1.2,1.05) ..
(X2);

\draw[black, line width=1.7pt]
(X1)
.. controls (-1.2,-1.05) and (1.2,-1.05) ..
(X2);

\fill[black] (X1) circle (4.8pt);
\fill[black] (X2) circle (4.8pt);

\end{tikzpicture}
}
}
&
11
&
121
&
84
\\
\hline


\raisebox{-0.5\height}{
\resizebox{!}{1.45cm}{
\begin{tikzpicture}[
    line cap=round,
    line join=round
]

\path[use as bounding box] (-4.8,-1.6) rectangle (4.8,3.0);

\coordinate (x1) at (-2.2,0);
\coordinate (x2) at ( 0.0,0);
\coordinate (x3) at ( 2.2,0);

\def\futureheight{2.6}

\draw[
    propagatorgray,
    line width=4pt
]
(-4.5,\futureheight) -- (4.5,\futureheight);

\draw[propagatorgray, line width=1.7pt]
    (x1) -- (-3.5,\futureheight);

\draw[propagatorgray, line width=1.7pt]
    (x1) -- (-2.8,\futureheight);

\draw[propagatorgray, line width=1.7pt]
    (x2) -- (-0.35,\futureheight);

\draw[propagatorgray, line width=1.7pt]
    (x2) -- (0.35,\futureheight);

\draw[propagatorgray, line width=1.7pt]
    (x3) -- (2.8,\futureheight);

\draw[propagatorgray, line width=1.7pt]
    (x3) -- (3.5,\futureheight);

\draw[black, line width=1.2pt]
(x1)
.. controls (-1.7,1.15) and (-0.5,1.15) ..
(x2);

\draw[black, line width=1.2pt]
(x1)
.. controls (-1.7,-1.15) and (-0.5,-1.15) ..
(x2);

\draw[black, line width=1.2pt]
    (x2) -- (x3);

\fill[black] (x1) circle (3.2pt);
\fill[black] (x2) circle (3.2pt);
\fill[black] (x3) circle (3.2pt);

\end{tikzpicture}
}
}
&
36
&
1\,296
&
1\,006
\\
\hline


\raisebox{-0.5\height}{
\resizebox{!}{1.45cm}{
\begin{tikzpicture}[
    line cap=round,
    line join=round
]

\path[use as bounding box] (-5.4,-2.1) rectangle (5.4,3.0);

\coordinate (X3) at (-4,0);
\coordinate (X4) at (-0.45,1.35);
\coordinate (X2) at ( 0.45,-1.35);
\coordinate (X1) at ( 4,0);

\def\futureheight{2.8}

\draw[
    propagatorgray,
    line width=4pt
]
(-5.0,\futureheight) -- (5.0,\futureheight);

\draw[propagatorgray, line width=1.7pt]
    (X3) -- (-4.7,\futureheight);

\draw[propagatorgray, line width=1.7pt]
    (X3) -- (-3.7,\futureheight);

\draw[propagatorgray, line width=1.7pt]
    (X4) -- (-1,\futureheight);

\draw[propagatorgray, line width=1.7pt]
    (X4) -- (-0.4,\futureheight);

\draw[propagatorgray, line width=1.7pt]
    (X2) -- (0.5,\futureheight);

\draw[propagatorgray, line width=1.7pt]
    (X2) -- (1.35,\futureheight);

\draw[propagatorgray, line width=1.7pt]
    (X1) -- (3.7,\futureheight);

\draw[propagatorgray, line width=1.7pt]
    (X1) -- (4.7,\futureheight);

\draw[black, line width=1.7pt] (X3) -- (X4);
\draw[black, line width=1.7pt] (X4) -- (X1);
\draw[black, line width=1.7pt] (X3) -- (X2);
\draw[black, line width=1.7pt] (X2) -- (X1);
\draw[black, line width=1.7pt] (X4) -- (X2);

\fill[black] (X1) circle (4.8pt);
\fill[black] (X2) circle (4.8pt);
\fill[black] (X3) circle (4.8pt);
\fill[black] (X4) circle (4.8pt);

\end{tikzpicture}
}
}
&
134
&
17\,956
&
15\,841
\\
\hline

\end{tabular}
\caption{Genealogical constraints for the bubble, chain-bubble, and acnode cosmological correlators. For each example, we report the number of singular loci, the corresponding number of ordered pairs of discontinuities that can na\"ively appear, and the number of these pairs that are forbidden by genealogical constraints.}
\label{fig:more_correlator_examples}
\end{table}

We have checked \DiscKosky{}'s predictions against further cosmological correlator examples, such as those shown in Figure~\ref{fig:additional_correlator_examples}. As each of these integrals involves many integrand terms, we derive genealogical constraints for each term separately and only report the constraints that are obeyed by all terms. The number of singular loci that appear in each example (which determines the number of ordered pairs of discontinuities that can na\"ively appear) is reported in Table~\ref{fig:more_correlator_examples}, alongside the number of hierarchical constraints~\eqref{eq:hierarchical-principle-background} that can be derived using the methods described above. It is encouraging to note that the constraining power of the hierarchical principle grows with the complexity of the integrals, to the point that over $88\%$ of the ordered pairs of discontinuities are forbidden from appearing in the most complicated example we consider. Note that all of these predictions are valid for general values of $T_i$ in~\eqref{eq:bubble_correlator}, as well as order by order in an expansion around $T_i=0$. In each of these examples, we have checked these predictions through (at least) the order in the $T_i=\epsilon\to0$ expansion at which weight five polylogarithms arise, by comparing to the results generated using~\cite{Larkin:2026ayy}.

%% file: Figures/bubble_correlator.tex
\begin{tikzpicture}[
    line cap=round,
    line join=round
]

\path[use as bounding box] (-5.2,-1.8) rectangle (5.2,3.0);


\definecolor{propagatorgray}{RGB}{202,202,218}


\coordinate (X1) at (-2.8,0);
\coordinate (X2) at ( 2.8,0);

\def\futureheight{2.6}


\draw[
    propagatorgray,
    line width=4pt
]
(-4.8,\futureheight) -- (4.8,\futureheight);


\draw[
    propagatorgray,
    line width=1.7pt
]
(X1) -- (-4.1,\futureheight);

\draw[
    propagatorgray,
    line width=1.7pt
]
(X1) -- (-3.3,\futureheight);

\draw[
    propagatorgray,
    line width=1.7pt
]
(X2) -- (3.3,\futureheight);

\draw[
    propagatorgray,
    line width=1.7pt
]
(X2) -- (4.1,\futureheight);


\draw[
    black,
    line width=1.7pt
]
(X1)
.. controls (-1.2,1.05) and (1.2,1.05) ..
(X2);

\draw[
    black,
    line width=1.7pt
]
(X1)
.. controls (-1.2,-1.05) and (1.2,-1.05) ..
(X2);


\fill[black] (X1) circle (4.8pt);
\fill[black] (X2) circle (4.8pt);


\node[
    font=\fontsize{13}{14}\selectfont,
    anchor=north east,
    xshift=-4pt,
    yshift=-4pt
]
at (X1)
{$X_1$};

\node[
    font=\fontsize{13}{14}\selectfont,
    anchor=north west,
    xshift=4pt,
    yshift=-4pt
]
at (X2)
{$X_2$};


\node[
    font=\fontsize{12}{13}\selectfont,
    fill=white,
    inner sep=1pt
]
at (0,0.72)
{$Y_1$};

\node[
    font=\fontsize{12}{13}\selectfont,
    fill=white,
    inner sep=1pt
]
at (0,-0.78)
{$Y_2$};

\end{tikzpicture}

%% file: Figures/bubble_critical_point_figure.tex
\begin{tikzpicture}[
    scale=0.88,
    transform shape,
    ->,
    thick,
    >={Triangle[length=1.5mm, width=2mm]}
]


\tikzset{
    sector/.style={
        draw,
        rounded corners=5pt,
        align=center,
        minimum width=1.55cm,
        minimum height=1.45cm,
        inner xsep=4pt,
        inner ysep=3pt
    },
    gene/.style={
        thick,
        shorten <=1pt,
        shorten >=1.5pt
    }
}


\def\rowone{0}
\def\rowtwo{-3.23}
\def\rowthree{-6.47}
\def\rowfour{-9.70}


\node (den1) at (-5.6,\rowone)   {1 denominator};
\node (den2) at (-5.6,\rowtwo)   {2 denominators};
\node (den3) at (-5.6,\rowthree) {3 denominators};
\node (den4) at (-5.6,\rowfour)  {4 denominators};


\node[sector, fill=blue!30] (0001) at (-2,0.14) {
    $\mathcal{D}_{0001}$\\[-8pt]
    \rule{1.35cm}{0.25pt}\\[-2pt]
    $C=1$\\
    $\Delta C=1$
};

\node[sector, fill=OliveGreen!70] (0010) at (2.5,-0.08) {
    $\mathcal{D}_{0010}$\\[-8pt]
    \rule{1.35cm}{0.25pt}\\[-2pt]
    $C=1$\\
    $\Delta C=1$
};

\node[sector] (0100) at (6.4,0.05) {
    $\mathcal{D}_{0100}$\\[-8pt]
    \rule{1.35cm}{0.25pt}\\[-2pt]
    $C=0$\\
    $\Delta C=0$
};

\node[sector] (1000) at (9.35,-0.16) {
    $\mathcal{D}_{1000}$\\[-8pt]
    \rule{1.35cm}{0.25pt}\\[-2pt]
    $C=0$\\
    $\Delta C=0$
};


\node[sector,fill=gray!10] (0011) at (-3.25,-2.84) {
    $\mathcal{D}_{0011}$\\[-8pt]
    \rule{1.35cm}{0.25pt}\\[-2pt]
    $C=2$\\
    $\Delta C=0$
};

\node[sector, fill=RedOrange!80] (0101) at (-0.55,-3.20) {
    $\mathcal{D}_{0101}$\\[-8pt]
    \rule{1.35cm}{0.25pt}\\[-2pt]
    $C=2$\\
    $\Delta C=1$
};

\node[sector, fill=Dandelion] (0110) at (2.15,-3.6) {
    $\mathcal{D}_{0110}$\\[-8pt]
    \rule{1.35cm}{0.25pt}\\[-2pt]
    $C=2$\\
    $\Delta C=1$
};

\node[sector, fill=brown!80] (1001) at (4.85,-3.1) {
    $\mathcal{D}_{1001}$\\[-8pt]
    \rule{1.35cm}{0.25pt}\\[-2pt]
    $C=2$\\
    $\Delta C=1$
};

\node[sector, fill=cyan!50] (1010) at (7.25,-2.88) {
    $\mathcal{D}_{1010}$\\[-8pt]
    \rule{1.35cm}{0.25pt}\\[-2pt]
    $C=2$\\
    $\Delta C=1$
};

\node[sector, fill=RedViolet!60] (1100) at (9.55,-3.25) {
    $\mathcal{D}_{1100}$\\[-8pt]
    \rule{1.35cm}{0.25pt}\\[-2pt]
    $C=1$\\
    $\Delta C=1$
};


\node[sector,fill=gray!10] (0111) at (-2.10,-6.20) {
    $\mathcal{D}_{0111}$\\[-8pt]
    \rule{1.35cm}{0.25pt}\\[-2pt]
    $C=4$\\
    $\Delta C=0$
};

\node[sector,fill=gray!10] (1011) at (2.00,-6.85) {
    $\mathcal{D}_{1011}$\\[-8pt]
    \rule{1.35cm}{0.25pt}\\[-2pt]
    $C=4$\\
    $\Delta C=0$
};

\node[sector,fill=gray!10] (1101) at (5.25,-6.45) {
    $\mathcal{D}_{1101}$\\[-8pt]
    \rule{1.35cm}{0.25pt}\\[-2pt]
    $C=4$\\
    $\Delta C=0$
};

\node[sector,fill=gray!10] (1110) at (8.60,-6.30) {
    $\mathcal{D}_{1110}$\\[-8pt]
    \rule{1.35cm}{0.25pt}\\[-2pt]
    $C=4$\\
    $\Delta C=0$
};


\node[sector] (1111) at (3.30,-9.70) {
    $\mathcal{D}_{1111}$\\[-8pt]
    \rule{1.35cm}{0.25pt}\\[-2pt]
    $C=7$\\
    $\Delta C=0$
};


\begin{scope}[on background layer]

    \node[
        fill=gray,
        fill opacity=0.1,
        rounded corners=8pt,
        minimum width=18cm,
        minimum height=2.80cm
    ] at (1.8,\rowone) {};

    \node[
        fill=gray,
        fill opacity=0.1,
        rounded corners=8pt,
        minimum width=18cm,
        minimum height=2.80cm
    ] at (1.8,\rowtwo) {};

    \node[
        fill=gray,
        fill opacity=0.1,
        rounded corners=8pt,
        minimum width=18cm,
        minimum height=2.80cm
    ] at (1.8,\rowthree) {};

    \node[
        fill=gray,
        fill opacity=0.1,
        rounded corners=8pt,
        minimum width=18cm,
        minimum height=2.80cm
    ] at (1.8,\rowfour) {};

\end{scope}


\begin{scope}[on background layer]


\draw[gene, color=blue!30]
    ([xshift=-18pt,yshift=1pt]0001.south)
    to[out=-105, in=90]
    (0011.north);

\draw[gene, color=blue!30]
    ([xshift=12pt,yshift=1pt]0001.south)
    to[out=-75, in=120]
    ([xshift=12pt]0101.north);

\draw[gene, color=blue!30]
    ([xshift=17pt,yshift=1pt]0001.south)
    to[out=-60,in=120]
    ([xshift=5pt]1001.north west);

\draw[gene, color=OliveGreen!70]
    ([xshift=-18pt,yshift=1pt]0010.south)
    to[out=-135, in=65]
    ([xshift=-4pt]0011.north east);
    
\draw[gene, color=OliveGreen!70]
    ([yshift=1pt,xshift=4pt]0010.south)
    to[out=-95, in=90]
    ([xshift=6pt]0110.north);

\draw[gene, color=OliveGreen!70]
    ([xshift=18pt,yshift=5pt]0010.south)
    to[out=-80,in=110]
    ([xshift=8pt]1010.north west);


\draw[gene, color=blue!30]
    ([xshift=-12pt,yshift=1pt]0001.south)
    to[out=-85, in=80]
    (0111.north);

\draw[gene, color=blue!30]
    ([yshift=1pt,xshift=0pt]0001.south)
    to[out=-80, in=105]
    ([xshift=-10pt]1011.north);

\draw[gene, color=blue!30]
    ([yshift=1pt,xshift=6pt]0001.south)
    to[out=-75, in=115]
    (1101.north);

\draw[gene, color=OliveGreen!70]
    ([yshift=1pt,xshift=-10pt]0010.south)
    to[out=-100,in=65]
    ([xshift=-5pt]0111.north east);

\draw[gene, color=OliveGreen!70]
    ([yshift=1pt,xshift=-5pt]0010.south)
    to[out=-100,in=90]
    ([xshift=-1pt]1011.north);

\draw[gene, color=OliveGreen!70]
    ([yshift=1pt,xshift=14pt]0010.south)
    to[out=-80, in=115]
    ([xshift=6pt]1110.north west);

\draw[gene, color=RedOrange!80]
    ([yshift=1pt,xshift=-10pt]0101.south)
    to[out=-115, in=75]
    ([xshift=8pt]0111.north);

\draw[gene, color=RedOrange!80]
    ([yshift=1pt,xshift=10pt]0101.south)
    to[out=-70,in=150]
    ([yshift=7pt]1101.west);

\draw[gene, color=Dandelion]
    ([yshift=1pt,xshift=-8pt]0110.south)
    to[out=-130,in=40]
    ([yshift=-10pt]0111.north east);

\draw[gene, color=Dandelion]
    ([yshift=1pt,xshift=14pt]0110.south)
    to[out=-65,in=130]
    ([yshift=-5pt]1110.north west);

\draw[gene, color=brown!80]
    ([yshift=1pt,xshift=7pt]1001.south)
    to[out=-90, in=90]
    ([xshift=8pt,]1101.north);

\draw[gene, color=brown!80]
    ([yshift=1pt,xshift=-8pt]1001.south)
    to[out=-120, in=75]
    ([xshift=13pt]1011.north);

\draw[gene, color=cyan!50]
    ([yshift=1pt,xshift=-10pt]1010.south)
    to[out=-110,in=55]
    ([yshift=-3pt]1011.north east);

\draw[gene, color=cyan!50]
    ([yshift=1pt,xshift=10pt]1010.south)
    to[out=-75, in=100]
    (1110.north);

\draw[gene, color=RedViolet!60]
    ([yshift=1pt,xshift=-10pt]1100.south)
    to[out=-110, in=55]
    (1101.north east);

\draw[gene, color=RedViolet!60]
    ([yshift=1pt,xshift=-3pt]1100.south)
    to[out=-95, in=80]
    ([xshift=9pt]1110.north);


\draw[gene, color=blue!30]
    ([yshift=1pt,xshift=-5pt]0001.south)
    to[out=-85, in=145]
    ([yshift=3pt]1111.west);

\draw[gene, color=OliveGreen!70]
    ([yshift=1pt,xshift=4pt]0010.south)
    to[out=90,in=90]
    ([xshift=4pt]1111.north);

\draw[gene, color=brown!80]
    ([yshift=1pt,xshift=-1pt]1001.south)
    to[out=-110, in=95]
    ([xshift=13pt]1111.north);

\draw[gene, color=cyan!50]
    ([yshift=1pt]1010.south)
    to[out=-90, in=35]
    ([yshift=-8pt]1111.north east);

\draw[gene, color=RedOrange!80]
    ([yshift=1pt,xshift=-2pt]0101.south)
    to[out=-85, in=140]
    ([yshift=-5pt]1111.north west);

\draw[gene, color=Dandelion]
    ([yshift=1pt,xshift=1pt]0110.south)
    to[out=-95,in=95]
    ([xshift=-8pt]1111.north);

\draw[gene, color=RedViolet!60]
    ([yshift=1pt,xshift=7pt]1100.south)
    to[out=-90,in=10]
    (1111.east);

\end{scope}

\end{tikzpicture}

%% file: Figures/cosmological_correlator_predictions_term_1.tex
\begin{tikzpicture}[%
    ->,
    shorten >=2pt,
    >=stealth,
    node distance=1cm,
    noname/.style={%
    color=Blue, minimum width=4em, minimum height=.1cm,scale=1}
  ]
    \node (title1) at (1.5,1.1) {Integrand Term 1};
    \node[draw, rounded corners=5pt,align=center] (1) at (0,0) {$\mathcal{D}_{0001}$\\\color{RedOrange}$\lambda_1$};
    \node[draw, rounded corners=5pt,align=center] (2) at (3,0) {$\mathcal{D}_{0010}$\\\color{RedOrange}$\lambda_2$};
    \node[draw, rounded corners=5pt,align=center] [below right=-0.5cm and 1cm of 2] (34) {$\mathcal{D}_{1100}$\\\color{RedOrange}$\lambda_8, \lambda_{10}$};
    \node[draw, rounded corners=5pt,align=center][below left=1cm and -.5cm of 1] (13) {$\mathcal{D}_{0101}$\\\color{RedOrange}$\lambda_7,\lambda_{10}$};
    \node[draw, rounded corners=5pt,align=center] [below right=1cm and -.5cm of 1] (14) {$\mathcal{D}_{1001}$\\\color{RedOrange}$\lambda_8,\lambda_{9}$};
    \node[draw, rounded corners=5pt,align=center] [below left=1cm and -.5cm of 2] (24) {$\mathcal{D}_{1010}$\\\color{RedOrange}$\lambda_5,\lambda_{8}$};
    \node[draw, rounded corners=5pt,align=center] [below right=1cm and -.5cm of 2] (23) {$\mathcal{D}_{0110}$\\\color{RedOrange}$\lambda_3,\lambda_{10}$};
    \path (1) edge [bend right=5pt, color=OliveGreen] node {}(13)
          (1) edge [bend left=10pt, color=OliveGreen] node {} (14)
          (2) edge [bend right=10pt, color=OliveGreen] node {} (24)
          (2) edge [bend left=10pt, color=OliveGreen] node {} (23);         
  \end{tikzpicture}

%% file: Figures/cosmological_correlator_predictions_term_2.tex
\begin{tikzpicture}[%
    ->,
    shorten >=2pt,
    >=stealth,
    node distance=1cm,
    noname/.style={%
    color=Blue, minimum width=4em, minimum height=.1cm,scale=1}
  ]
    \node (title2) at (1.5,1.1) {Integrand Term 2};
    \node[draw, rounded corners=5pt,align=center] (1) at (0,0) {$\mathcal{D}^\prime_{0001}$\\\color{RoyalBlue}$\lambda_1$};
    \node[draw, rounded corners=5pt,align=center] (2) at (3,0) {$\mathcal{D}^\prime_{0010}$\\\color{RoyalBlue}$\lambda_{11}$};
    \node[draw, rounded corners=5pt,align=center] [below right=-0.5cm and 1cm of 2] (34) {$\mathcal{D}^\prime_{1100}$\\\color{RoyalBlue}$\lambda_8, \lambda_{10}$};
    \node[draw, rounded corners=5pt,align=center][below left=1cm and -.5cm of 1] (13) {$\mathcal{D}^\prime_{0101}$\\\color{RoyalBlue}$\lambda_7,\lambda_{10}$};
    \node[draw, rounded corners=5pt,align=center] [below right=1cm and -0.5cm of 1] (14) {$\mathcal{D}^\prime_{1001}$\\\color{RoyalBlue}$\lambda_8,\lambda_{9}$};
    \node[draw, rounded corners=5pt,align=center] [below left=1cm and -0.5cm of 2] (24) {$\mathcal{D}^\prime_{1010}$\\\color{RoyalBlue}$\lambda_6,\lambda_{8}$};
    \node[draw, rounded corners=5pt,align=center] [below right=1cm and -.5cm of 2] (23) {$\mathcal{D}^\prime_{0110}$\\\color{RoyalBlue}$\lambda_4,\lambda_{10}$};
    \path (1) edge [bend right=5pt, color=OliveGreen] node {}(13)
          (1) edge [bend left=10pt, color=OliveGreen] node {} (14)
          (2) edge [bend right=10pt, color=OliveGreen] node {} (24)
          (2) edge [bend left=10pt, color=OliveGreen] node {} (23);         
  \end{tikzpicture}

%% file: Figures/ball_chain_correlator.tex
\begin{tikzpicture}[
    line cap=round,
    line join=round
]

\path[use as bounding box] (-4.8,-1.6) rectangle (4.8,3.0);


\definecolor{propagatorgray}{RGB}{202,202,218}


\coordinate (x1) at (-2.2,0);
\coordinate (x2) at ( 0.0,0);
\coordinate (x3) at ( 2.2,0);

\def\futureheight{2.6}


\draw[
    propagatorgray,
    line width=4pt
]
(-4.5,\futureheight) -- (4.5,\futureheight);


\draw[
    propagatorgray,
    line width=1.7pt
]
(x1) -- (-3.5,\futureheight);

\draw[
    propagatorgray,
    line width=1.7pt
]
(x1) -- (-2.8,\futureheight);

\draw[
    propagatorgray,
    line width=1.7pt
]
(x2) -- (-0.35,\futureheight);

\draw[
    propagatorgray,
    line width=1.7pt
]
(x2) -- (0.35,\futureheight);

\draw[
    propagatorgray,
    line width=1.7pt
]
(x3) -- (2.8,\futureheight);

\draw[
    propagatorgray,
    line width=1.7pt
]
(x3) -- (3.5,\futureheight);


\draw[
    black,
    line width=1.2pt
]
(x1)
.. controls (-1.7,1.15) and (-0.5,1.15) ..
(x2);

\draw[
    black,
    line width=1.2pt
]
(x1)
.. controls (-1.7,-1.15) and (-0.5,-1.15) ..
(x2);


\draw[
    black,
    line width=1.2pt
]
(x2) -- (x3);


\fill[black] (x1) circle (3.2pt);
\fill[black] (x2) circle (3.2pt);
\fill[black] (x3) circle (3.2pt);


\node[
    font=\fontsize{13}{14}\selectfont,
    anchor=east,
    xshift=-4pt
]
at (x1)
{$X_1$};

\node[
    font=\fontsize{13}{14}\selectfont,
    anchor=south west,
    xshift=2pt,
    yshift=4pt
]
at (x2)
{$X_2$};

\node[
    font=\fontsize{13}{14}\selectfont,
    anchor=west,
    xshift=4pt
]
at (x3)
{$X_3$};


\node[
    font=\fontsize{12}{13}\selectfont,
    fill=white,
    inner xsep=2pt,
    inner ysep=1pt
]
at (-1.1,0.78)
{$Y_1$};

\node[
    font=\fontsize{12}{13}\selectfont,
    fill=white,
    inner xsep=2pt,
    inner ysep=1pt
]
at (-1.1,-0.81)
{$Y_2$};

\node[
    font=\fontsize{12}{13}\selectfont,
    fill=white,
    inner xsep=2pt,
    inner ysep=1pt
]
at (1.12,-0)
{$Y_3$};

\end{tikzpicture}

%% file: Figures/acnode_correlator.tex
\begin{tikzpicture}[
    line cap=round,
    line join=round
]

\path[use as bounding box] (-5.4,-2.1) rectangle (5.4,3.0);


\definecolor{propagatorgray}{RGB}{202,202,218}


\coordinate (X3) at (-4,0);
\coordinate (X4) at (-0.45,1.35);
\coordinate (X2) at ( 0.45,-1.35);
\coordinate (X1) at ( 4,0);

\def\futureheight{2.8}


\draw[
    propagatorgray,
    line width=4pt
]
(-5.0,\futureheight) -- (5.0,\futureheight);


\draw[propagatorgray, line width=1.7pt]
    (X3) -- (-4.7,\futureheight);

\draw[propagatorgray, line width=1.7pt]
    (X3) -- (-3.7,\futureheight);

\draw[propagatorgray, line width=1.7pt]
    (X4) -- (-1,\futureheight);

\draw[propagatorgray, line width=1.7pt]
    (X4) -- (-0.4,\futureheight);

\draw[propagatorgray, line width=1.7pt]
    (X2) -- (0.5,\futureheight);

\draw[propagatorgray, line width=1.7pt]
    (X2) -- (1.35,\futureheight);

\draw[propagatorgray, line width=1.7pt]
    (X1) -- (3.7,\futureheight);

\draw[propagatorgray, line width=1.7pt]
    (X1) -- (4.7,\futureheight);


\draw[black, line width=1.7pt] (X3) -- (X4); 
\draw[black, line width=1.7pt] (X4) -- (X1); 
\draw[black, line width=1.7pt] (X3) -- (X2); 
\draw[black, line width=1.7pt] (X2) -- (X1); 
\draw[black, line width=1.7pt] (X4) -- (X2); 


\fill[black] (X1) circle (4.8pt);
\fill[black] (X2) circle (4.8pt);
\fill[black] (X3) circle (4.8pt);
\fill[black] (X4) circle (4.8pt);


\node[
    font=\fontsize{13}{14}\selectfont,
    anchor=south east,
    xshift=-3pt,
    yshift=2pt
]
at (X3)
{$X_3$};

\node[
    font=\fontsize{13}{14}\selectfont,
    anchor=south west,
    xshift=3pt,
    yshift=2pt
]
at (X1)
{$X_1$};

\node[
    font=\fontsize{13}{14}\selectfont,
    anchor=north east,
    xshift=1pt,
    yshift=-6pt
]
at (X4)
{$X_4$};

\node[
    font=\fontsize{13}{14}\selectfont,
    anchor=north,
    yshift=-5pt
]
at (X2)
{$X_2$};


\node[
    font=\fontsize{12}{13}\selectfont,
    fill=white,
    inner xsep=1.8pt,
    inner ysep=1.3pt
]
at (-2.0,0.72)
{$Y_3$};

\node[
    font=\fontsize{12}{13}\selectfont,
    fill=white,
    inner xsep=1.8pt,
    inner ysep=1.3pt
]
at (1.75,0.7)
{$Y_4$};

\node[
    font=\fontsize{12}{13}\selectfont,
    fill=white,
    inner xsep=1.8pt,
    inner ysep=1.3pt
]
at (-1.85,-0.72)
{$Y_2$};

\node[
    font=\fontsize{12}{13}\selectfont,
    fill=white,
    inner xsep=1.8pt,
    inner ysep=1.3pt
]
at (1.80,-0.8)
{$Y_1$};

\node[
    font=\fontsize{12}{13}\selectfont,
    fill=white,
    inner xsep=2pt,
    inner ysep=3pt
]
at (0.02,0)
{$Y_5$};

\end{tikzpicture}

%% file: sections/non_repeating_letters.tex
\section{Non-Repeating Discontinuities}\label{sec:non_repeating}

In this section, we explore a previously overlooked consequence of the hierarchical principle---that some discontinuities are forbidden from appearing in Feynman integrals more than once. More specifically, we identify discontinuities that, once computed, give rise to integrals in which the on-shell configuration responsible for the original discontinuity can no longer be realized. To elucidate how this behavior arises, we study it in three examples of increasing complexity. We also formulate a general test that allows us to identify such discontinuities using the critical-point-counting methods developed in Section~\ref{sect:theoretical-background}, which we have implemented as part of \DiscKosky{}.

\subsection{Example 1: the Hook Triangle}

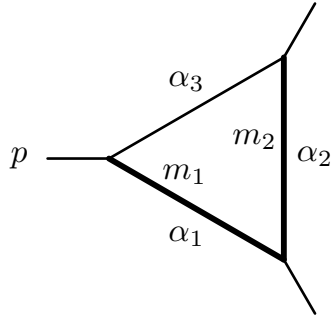
\begin{figure}[t]
    \centering
\begin{tikzpicture}[
    scale=1.25,
    transform shape,
    line cap=round,
    line join=round,
    ext/.style={line width=1pt},
    prop/.style={line width=1pt},
    heavy/.style={line width=2.2pt}
]

    \coordinate (A) at (-1.1,0);
    \coordinate (B) at (0.75,1.07);
    \coordinate (C) at (0.75,-1.07);

    \coordinate (Pin) at (-1.75,0);
    \coordinate (Pup) at (1.08,1.64);
    \coordinate (Pdn) at (1.08,-1.64);

    \draw[prop]  (A) -- (B);
    \draw[heavy] (A) -- (C);
    \draw[heavy] (C) -- (B);

    \draw[ext] (Pin) -- (A);
    \draw[ext] (B) -- (Pup);
    \draw[ext] (C) -- (Pdn);

    \node at (-2.05,0) {\small\textit{$p$}};
    \node at (-0.28,0.82) {\small\textit{$\alpha_3$}};
    \node at (-0.28,-0.82) {\small\textit{$\alpha_1$}};
    \node at (1.08,0.02) {\small\textit{$\alpha_2$}};

    \node at (-0.3,-0.17) {\small\textit{$m_1$}};
    \node at (0.44,0.20) {\small\textit{$m_2$}};

\end{tikzpicture}
\caption{The one-loop hook triangle integral, with internal masses $m_1$ and $m_2$, and propagators labeled by the corresponding Feynman parameters $\alpha_1$, $\alpha_2$, and $\alpha_3$.}    \label{fig:hook triange}
\end{figure}

One of the simplest examples in which this phenomenon occurs is the hook triangle diagram, shown in Figure~\ref{fig:hook triange}. Already in~\cite{britto2025trackingdiscontinuitiesparameterspace}, this integral was found to exhibit interesting behavior with respect to the singularity at $m_1^2-m_2^2=0$. Namely, although a logarithmic discontinuity arises at this locus after one leaves the principal Riemann sheet (as can be seen from the fact that $\text{Disc}_{m_1^2-m_2^2}\,\text{Disc}_{m_1^2}\, \mathcal{I}_{\hooktriangle}\neq 0$ and $\text{Disc}_{m_1^2-m_2^2}\,\text{Disc}_{m_2^2}\,\mathcal{I}_{\hooktriangle}\neq 0$), immediately computing $\text{Disc}_{m_1^2-m_2^2}$ a second time annihilates the integral:
\begin{equation}
    \text{Disc}_{m_1^2-m_2^2}\,
    \text{Disc}_{m_1^2-m_2^2}\,
    \text{Disc}_{m_1^2}\,
    \mathcal{I}_{\hooktriangle}
    =0\, .
\end{equation}
Note that this statement holds to all orders in dimensional regularization, so it does not merely reflect the transcendental weight of the hook triangle when considered in exactly four dimensions.

As it turns out, an even stronger result holds. By inspecting the canonical differential equation satisfied by this integral, one finds that, after a single discontinuity with respect to $m_1^2-m_2^2=0$ has been computed, no further discontinuity can arise on the same kinematic hypersurface (regardless of whether additional analytic continuations are carried out). This places strong constraints on the symbol of the integral to all orders in dimensional regularization. 

This behavior has a simple interpretation in terms of the critical-point equations. Using the methods of Section~\ref{sect:theoretical-background}, the hook triangle integral is found to have four critical points, each of which lies in a different sector:
\begin{equation}
    S(\sigma_1)=\{1\},\qquad
    S(\sigma_2)=\{2\},\qquad
    S(\sigma_3)=\{1,3\},\qquad
    S(\sigma_4)=\{1,2,3\}.
    \label{eq:hook-triangle-sector-data}
\end{equation}
Surprisingly, none of these critical points vanish when we set $m_1^2-m_2^2=0$. Rather, in the strict $m_1^2-m_2^2\to0$ limit, a one-dimensional family of solutions appears in the $\{1,2\}$ sector. Whilst this doesn't correspond to a drop in the number of critical points in the way described in Section~\ref{sect:theoretical-background}, this discontinuous change nevertheless signals a change in topology of the integration space. We thus associate the $m_1^2-m_2^2=0$ singularity with the $\{1,2\}$ sector.

We now ask what the integration contour will look like after we have computed a discontinuity with respect to $m_1^2-m_2^2=0$. Recall that each of the four critical points~\eqref{eq:hook-triangle-sector-data} defines a Lefschetz thimble (or steepest descent contour), and that together these thimbles form a basis of integration cycles that can appear in the hook triangle integral. As such, $\text{Disc}_{m_1^2-m_2^2} \cdots \mathcal{I}_{\hooktriangle}$ must correspond to an integral over some linear combination of these cycles. However, because the $m_1^2-m_2^2=0$ singularity lives in sector $\{1,2\}$, propagators $1$ and $2$ must both be on shell after this discontinuity has been computed---and of the four critical points, only $\sigma_4$ still exists once these propagators are on shell. It follows that the thimble associated with $\sigma_4$ must constitute the (unique) vanishing cycle for the $m_1^2-m_2^2=0$ discontinuity. 

Crucially, since the singularity at $m_1^2-m_2^2=0$ lives in sector $\{1,2\}$, it can no longer be reached in sector $\{1,2,3\}$. Thus, once a first discontinuity with respect to the singularity at $m_1^2-m_2^2=0$ has been computed, no further branch cuts can arise at $m_1^2-m_2^2=0$. Indeed, the only singularities that can be reached in the $\{1,2,3\}$ sector are
\begin{equation}
    p^2=0,\qquad
    m_2^2=0,\qquad
    p^2+m_2^2-m_1^2=0\, .
\end{equation}
It can be checked that these are precisely the kinematic loci where discontinuities still arise after $\text{Disc}_{m_1^2-m_2^2}$ has been computed.

\subsection{Example 2: the Two-Mass Easy Box}

A second example in which non-repeating discontinuities arise is the two-mass easy box integral $\mathcal{I}_{\twomasseasybox}$, which we already encountered in Section~\ref{sec:quickstart} and which was depicted in Figure~\ref{fig:2m-easy}. In generic kinematics, this integral has five critical points that live in the sectors
\begin{equation}
\begin{gathered}
    S(\sigma_1)=\{1,4\},\qquad
    S(\sigma_2)=\{2,3\},\qquad
    S(\sigma_3)=\{1,3\},\qquad
    S(\sigma_4)=\{2,4\},\\
    S(\sigma_5)=\{1,2,3,4\}.
\end{gathered}
\label{eq:ci-sets}
\end{equation}
Its symbol involves 10 letters, and in~\cite{Hannesdottir:2024cnn} it was shown that hierarchical constraints forbid 64 ordered pairs of (distinct) letters from appearing in each symbol term. We now show that four additional constraints can be derived, each of which forbids an individual letter from appearing more than once in a given term.

We carry out the same type of analysis as for the hook triangle, focusing first on the $s_{12}-s_{123}=0$ hypersurface. In the $s_{12}-s_{123}\to0$ limit, none of the five critical points in~\eqref{eq:ci-sets} vanish, but a new one-dimensional solution to the critical point equations appears in sector $\{1,2,4\}$. Following the same reasoning as in the hook triangle example, the integration cycle that appears in $\text{Disc}_{s_{12}-s_{123}} \cdots \mathcal{I}_{\twomasseasybox}$ must be drawn from sectors in which propagators 1, 2, and 4 are all on shell. The sole candidate is the thimble associated with  $\sigma_5$, which lives in sector $\{1,2,3,4\}$. However, as this integration contour is localized to the on-shell locus of all four propagators, the $s_{12}-s_{123}=0$ singularity can no longer be reached. It follows that $s_{12}-s_{123}$ can no longer appear as a symbol letter.

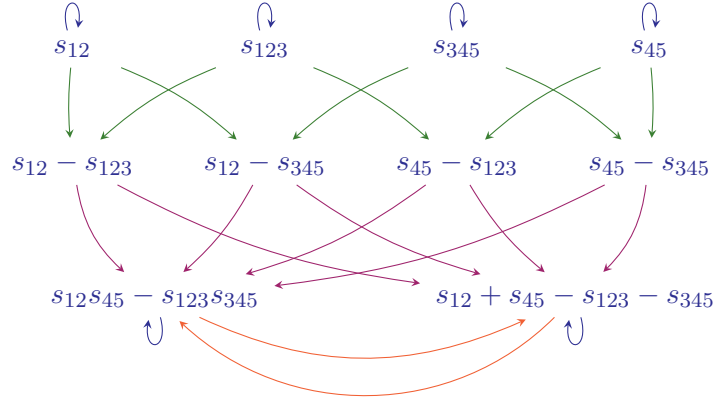
\begin{figure}\centering
\input{Figures/two_mass_easy_hierarchical}
  \caption{The sequences of discontinuities that are allowed to appear in the two-mass easy box integral. An arrow $a \to b$ indicates that there can exist a singularity at $b = 0$ after a discontinuity has been computed with respect to $a = 0$. Note that the four letters in the middle row do not have arrows pointing back to themselves; this indicates that these singularities correspond to non-repeating discontinuities.}
  \label{fig:two_mass_easy_hierarchical}
\end{figure}

The same argument can be carried out for the letters $s_{12}-s_{612}$, $s_{45}-s_{345}$, and $s_{45}-s_{456}$, which are all related to $s_{12}-s_{123}$ by symmetry. In each case, the singularity appears in a sector involving three propagators, implying that the discontinuity with respect to each of these singularities involves the unique integration cycle that lives in sector $\{1,2,3,4\}$.\footnote{Note that this observation has the additional consequence that $\text{Disc}_{s_{12}-s_{123}} \cdots \mathcal{I}_{\twomasseasybox}$, $\text{Disc}_{s_{12}-s_{612}} \cdots \mathcal{I}_{\twomasseasybox}$, $\text{Disc}_{s_{45}-s_{345}} \cdots \mathcal{I}_{\twomasseasybox}$, and $\text{Disc}_{s_{45}-s_{456}} \cdots \mathcal{I}_{\twomasseasybox}$ must all evaluate to the same function, up to a numerical proportionality factor (no matter what initial sequence of discontinuities has first been computed), as explained in~\cite{McLeod:2026kaf}.} As such, none of these letters can be repeated in the symbol. These results, together with the other genealogical constraints derived in~\cite{Hannesdottir:2024cnn}, are summarized in Figure~\ref{fig:two_mass_easy_hierarchical}.\footnote{Note that this way of representing which sequences of discontinuities are allowed by the hierarchical principle is slightly different than the graphs drawn by \DiscKosky{}, as illustrated in~\eqref{eq:disckosky_graph}. In Figure~\ref{fig:two_mass_easy_hierarchical}, the nodes of the graph correspond to the singularities of the integral, while the nodes of the \DiscKosky{} graphs represent different sectors. Ultimately, these different ways of representing hierarchical constraints encode the same data.}

\subsection{The Non-Repeating Discontinuity Test}

The two examples above suggest a simple test for identifying non-repeating discontinuities. Suppose that we have associated the singular hypersurface $\lambda=0$ with one or more sectors $S_{\lambda,i}$ by identifying either a drop in the number of critical points or the emergence of a new higher-dimensional solution to the critical point equations in the $\lambda \to 0$ limit (as observed in the hook triangle and two-mass easy box examples). For the sake of simplicity, let us assume there is just one singular sector $S_\lambda$. After $\text{Disc}_{\lambda}$ has been computed, the hierarchical principle then restricts the integration contour to (linear combinations of) thimbles whose sectors involve (at least) all the same on-shell conditions as $S_\lambda$. If a generic critical point already exists in this sector, we reproduce the standard hierarchical constraints discussed in Section~\ref{sect:theoretical-background}. However, if $S_{\lambda}$ contains no critical point in generic kinematics, there is no thimble associated directly with this sector, and the discontinuity must instead be supported on thimbles associated with critical points $\sigma$ that satisfy
\begin{equation} \label{eq:sectors-supporting-discontinuity}
    S_{\lambda}\subseteq S(\sigma)\, .
\end{equation}
Computing a discontinuity with respect to $\lambda=0$ thus forces the integration contour into sectors in which additional propagators are put on shell. Since, by assumption, $\lambda=0$ is associated only with $S_\lambda$, the same discontinuity can never again occur.\footnote{The same reasoning can also be used to constrain sequences of discontinuities involving distinct singularities. For instance, if $\lambda=0$ and $\lambda^\prime=0$ arise (only) in the same sector, and $\lambda=0$ corresponds to a non-repeating discontinuity, then we can also deduce that $\text{Disc}_{\lambda^\prime}\cdots\text{Disc}_{\lambda}\cdots\mathcal{I}=0$.}

This test is implemented in \DiscKosky{} through the \texttt{"RefineIndeterminates"} option described in Section~\ref{sec:core_usage}. When \texttt{"RefineIndeterminates"} is enabled, \DiscKosky{} identifies sectors in which no critical points appear in generic kinematics but that develop a higher-dimensional solution to the critical point equations when $\lambda=0$. In such cases, the code then enumerates the generic critical points $\sigma$ whose sectors satisfy~\eqref{eq:sectors-supporting-discontinuity}, and applies the hierarchical principle to determine which discontinuities can follow $\text{Disc}_{\lambda}$ from its knowledge of which singularities appear in these sectors. If several generic critical points satisfy~\eqref{eq:sectors-supporting-discontinuity}, all of the corresponding sectors are retained, so that the resulting predictions remain conservative. Using the \texttt{"RefineIndeterminates"} option, \DiscKosky{} correctly identifies the non-repeating discontinuities in the hook triangle and the two-mass easy box, and provides an automated way of detecting the same behavior in more complicated Feynman integrals.

\subsection{Example 3: the One-Mass Five-Point Double Box}\label{sect:hexabox}

\begin{figure}
\centering
\begin{tikzpicture}[
    scale=1.25,
    line cap=round,
    line join=round,
    ext/.style={line width=1pt},
    prop/.style={line width=1pt},
    massive/.style={line width=2.2pt}
]

    \coordinate (n1) at (0,0.75);
    \coordinate (n2) at (1.25,0.75);
    \coordinate (n3) at (1.25,-0.75);
    \coordinate (n4) at (0,-0.75);
    \coordinate (n5) at (-0.75,0);

    \coordinate (c) at (0.625,0);

    \coordinate (g1) at ($(n1)!0.45!(n3)$);
    \coordinate (g2) at ($(n1)!0.55!(n3)$);

    \coordinate (p1) at (-0.38,1.18);
    \coordinate (p2) at (1.63,1.18);
    \coordinate (p3) at (-0.38,-1.18);
    \coordinate (p4) at (-1.18,0);
    \coordinate (p5) at (1.63,-1.18);

    \draw[prop] (n1) -- (n2);
    \draw[prop] (n2) -- (n4);
    \draw[prop] (n4) -- (n5) -- (n1);
    \draw[prop] (n4) -- (n3);

    \draw[prop] (n1) -- (g1);
    \draw[prop] (g2) -- (n3);

    \draw[massive] (n1) -- (p1);
    \draw[ext]     (n2) -- (p2);
    \draw[ext]     (n4) -- (p3);
    \draw[ext]     (n5) -- (p4);
    \draw[ext]     (n3) -- (p5);

    \node[above left=2pt]  at (p1) {$p_1$};
    \node[above right=2pt] at (p2) {$p_2$};
    \node[below left=2pt]  at (p3) {$p_3$};
    \node[left=3pt]        at (p4) {$p_4$};
    \node[below right=2pt] at (p5) {$p_5$};

\end{tikzpicture}
\caption{A one-mass five-point double box integral. The massive external momentum $p_1$ is indicated by the bold external line.}
\label{fig:5pt 1-mass np}
\end{figure}
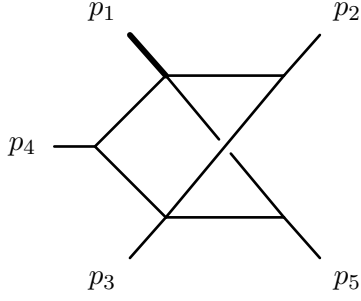

With this new test in hand, let us consider one final example, the one-mass five-point double box shown in Figure~\ref{fig:5pt 1-mass np}. The symbol alphabet of this integral involves 22 rational letters and 7 algebraic letters, and in~\cite{Hannesdottir:2024cnn} it was shown that genealogical constraints forbid 540 ordered pairs of these letters from appearing.\footnote{Note that this number includes restrictions involving algebraic letters. We discuss constraints on algebraic letters in more detail in the next section.} Running \DiscKosky{} reproduces these constraints. However, when \texttt{"RefineIndeterminates"} is set to \texttt{True}, the code finds an additional 9 non-repeating discontinuity constraints:
\begin{gather}
      \text{Disc}_{\lambda} \cdots \text{Disc}_{\lambda} \cdots \mathcal{I}_{\dblbox} = 0\, , \\[.1cm]
      \begin{aligned}
\lambda \in \big\{&
p_1^2{-}s_{12},\
p_1^2{-}s_{51},\
s_{12}{-}s_{34},\
s_{34}{+}s_{45},\
s_{51}{-}s_{23},\
s_{51}{-}s_{34},\\
& \qquad \qquad 
s_{51}{-}s_{23}{+}s_{45},\
p_1^2{-}s_{51}{-}s_{45},\ p_1^2{-}s_{12}{-}s_{51}{+}s_{23}{+}s_{34}
\big\}\, .
\end{aligned}
\end{gather}
These are precisely the 9 (empirical) hierarchical constraints that~\cite{Hannesdottir:2024cnn} reported not being able to derive. Thus, in this case, we now recover the implications of the hierarchical principle, as checked against the canonical differential equation derived in~\cite{Abreu:2021smk}.

\subsection{The Nature of Non-Repeating Discontinuities}

The analysis above also tells us something about the type of singularity that can give rise to a non-repeating discontinuity. Suppose that a singularity at $\lambda=0$ arises in a single sector, and that we have identified it as giving rise to a non-repeating discontinuity. Then one can show that this singularity cannot give rise to an algebraic branch cut. 

Suppose, for instance, that our integral gave rise to a square-root branch point at $\lambda=0$. We can then decompose the full integral $\mathcal{F}(\sqrt{\lambda})$ into parts that are even and odd under the exchange of the two square-root branches:
\begin{equation}
    \mathcal{F}(\sqrt{\lambda})
    =
    \mathcal{F}_{+}(\sqrt{\lambda})
    +
    \mathcal{F}_{-}(\sqrt{\lambda})\, ,
    \qquad
    \mathcal{F}_{\pm}(-\sqrt{\lambda})
    =
    \pm\mathcal{F}_{\pm}(\sqrt{\lambda})\, .
\end{equation}
The presence of a nontrivial square-root branch point implies that $\mathcal{F}_{-}\neq0$. Thus, analytically continuing around $\lambda=0$ to compute a discontinuity with respect to this branch point, we get
\begin{equation}
    \text{Disc}_{\lambda}\mathcal{F}
    =
    \mathcal{F}(-\sqrt{\lambda})
    -
    \mathcal{F}(\sqrt{\lambda})
    =
    -2\mathcal{F}_{-}(\sqrt{\lambda})\, .
\end{equation}
Computing a second discontinuity with respect to $\lambda=0$ then gives
\begin{equation}
    \text{Disc}_{\lambda}^{\,2}\mathcal{F}
    =
    4\mathcal{F}_{-}(\sqrt{\lambda})\, .
\end{equation}
Thus, whenever a genuine square-root branch point contributes a nonzero discontinuity, the same discontinuity can always be computed again. This contradicts the assumption that $\lambda=0$ gives rise to a non-repeating discontinuity, so we see that a square-root branch point cannot arise at this point.

Similar arguments can be used to rule out any other non-integer power of $\lambda$ from appearing in the final result. We conclude that any nontrivial branch cut associated with a non-repeating discontinuity in Feynman integrals must be logarithmic in nature.

%% file: Figures/two_mass_easy_hierarchical.tex
\begin{tikzpicture}[%
    ->,
    shorten >=2pt,
    >=stealth,
    node distance=1cm,
    noname/.style={%
    color=Blue, minimum width=4em, minimum height=.1cm,scale=1}
  ]
    \node[noname] (12) {$s_{12}$};
    \node[noname] (123) [right=of 12] {$s_{123}$};
    \node[noname] (345) [right=of 123] {$s_{345}$};
    \node[noname] (45) [right=of 345] {$s_{45}$};
    \node[noname] (12m123) [below=of 12] {$s_{12} - s_{123}$};
    \node[noname] (12m345) [below=of 123] {$s_{12} - s_{345}$};
    \node[noname] (45m123) [below=of 345] {$s_{45} - s_{123}$};
    \node[noname] (45m345) [below=of 45] {$s_{45} - s_{345}$};
    \node[noname] (12t45m123t345) [node distance=1.2cm and -1cm,below left=of 12m345] {$s_{12} s_{45} - s_{123}s_{345}$};
    \node[noname] (12m123m345p45) [node distance=1.2cm and -1.36cm,below right=of 45m123] {$s_{12} + s_{45} - s_{123} - s_{345}$};
        \path (12) edge [bend right=5pt, color=OliveGreen] node {}(12m123)
          (12) edge [bend left=10pt, color=OliveGreen] node {} (12m345)
          (123) edge [bend right=10pt, color=OliveGreen] node {} (12m123)
          (123) edge [bend left=10pt, color=OliveGreen] node {} (45m123)         
          (345) edge [bend right=10pt, color=OliveGreen] node {} (12m345)
          (345) edge [bend left=10pt, color=OliveGreen] node {} (45m345)
          (45) edge [bend right=10pt, color=OliveGreen] node {} (45m123)
          (45) edge [bend left=5pt, color=OliveGreen] node {} (45m345)
          (12m123) edge [bend right=20pt, color=RedViolet] node {} (12t45m123t345)
          (12m123) edge [bend right=10pt, color=RedViolet] node {} (12m123m345p45)    
          (12m345) edge [bend left=10pt, color=RedViolet] node {} (12t45m123t345)
          (12m345) edge [bend right=10pt, color=RedViolet] node {} (12m123m345p45)
          (45m123) edge [bend left=10pt, color=RedViolet] node {} (12t45m123t345)
          (45m123) edge [bend right=10pt, color=RedViolet] node {} (12m123m345p45)
          (45m345) edge [bend left=10pt, color=RedViolet] node {} (12t45m123t345)
          (45m345) edge [bend left=20pt, color=RedViolet] node {} (12m123m345p45)
          (12t45m123t345) edge [bend right=25pt, color=RedOrange] node {} (12m123m345p45)
          (12m123m345p45) edge [bend left=45pt, color=RedOrange] node {} (12t45m123t345);
    \path (12) edge [loop above, color=Blue] node {} (12)
          (123) edge [loop above, color=Blue] node {} (123)
          (345) edge [loop above, color=Blue] node {} (345)
          (45) edge [loop above, color=Blue] node {} (45)
          (12t45m123t345) edge [loop below, color=Blue] node {} (12t45m123t345)
          (12m123m345p45) edge [loop below, color=Blue] node {} (12m123m345p45);
    \end{tikzpicture}

%% file: sections/genelogical_on_individual_symbol.tex
\section{Genealogical Constraints and Algebraic Symbol Letters}
\label{sec:algebraic-letters}

In the previous section, we saw that non-repeating discontinuities cannot be associated with algebraic branch points. Nevertheless, square roots frequently arise in Feynman integrals, and the hierarchical principle---which applies to all types of discontinuities---places constraints on how these square roots can appear together with logarithmic branch points. In particular, a single algebraic symbol letter can encode several branch
points, each of which can place constraints on how this letter appears in
the final result.

To see how these constraints manifest, consider an algebraic letter of the form
\begin{equation}
    A=\frac{a+\sqrt{b}}{a-\sqrt{b}}\, .
    \label{eq:algebraic-letter-generic}
\end{equation}
This letter has an algebraic branch point at $b=0$, since analytically continuing around this locus exchanges the two branches of the square root and sends $A\to 1/A$. At the same time, $A$ can give rise to logarithmic branch points whenever its numerator or denominator vanishes. Such loci are contained in the hypersurface defined by
\begin{equation}
    \left( a+\sqrt{b} \right) \left( a-\sqrt{b} \right)  = a^2-b=0\, .
    \label{eq:algebraic-letter-log-loci}
\end{equation}
Thus, if $\lambda$ is an irreducible factor of $a^2-b$ and $A$ vanishes or diverges as $\lambda\to0$, the same symbol letter encodes both the algebraic singularity at $b=0$ and a logarithmic singularity at $\lambda=0$.

Already by itself, the letter \(A\) therefore encodes nontrivial sequences of discontinuities. Indeed, computing a discontinuity with respect to \(b=0\) gives
\begin{align}
    \text{Disc}_{b}\log A
    &=
    \log \frac{1}{A}-\log A \\
    &=
    -2\log A+\mathcal{O}(2\pi i)\, .
\end{align}
One can therefore compute a subsequent discontinuity with respect to any of the logarithmic branch points \(\lambda=0\) that appear in \(A\). At first sight, this seems to imply that the letter \(A\) can appear only if a discontinuity with respect to \(\lambda=0\) is hierarchically allowed to follow a discontinuity with respect to \(b=0\). However, additional dependence on \(\sqrt{b}\) can hide the algebraic branch cut. For example, the function
\begin{equation}
    \mathcal{F}
    =
    \frac{1}{\sqrt{b}}\log A
\end{equation}
has no discontinuity at \(b=0\) on its principal sheet. Rather, the branch cut at $b=0$ is only exposed after a discontinuity has been computed with respect to one of the logarithmic branch points (or more generally, after one has analytically continued around one of these logarithmic branch points, generating an additional contribution proportional to $i \pi /\sqrt{b}$). Thus, in this example, we have that
\begin{equation}
    \text{Disc}_{\lambda}\,
    \text{Disc}_{b}\,\mathcal{F}=0\, ,
    \qquad
    \text{Disc}_{b}\,
    \text{Disc}_{\lambda}\,\mathcal{F}\neq0\, .
\end{equation}
Correspondingly, we see that hierarchical constraints cannot be applied to individual symbol letters without knowing more about the global structure of the Feynman integral.

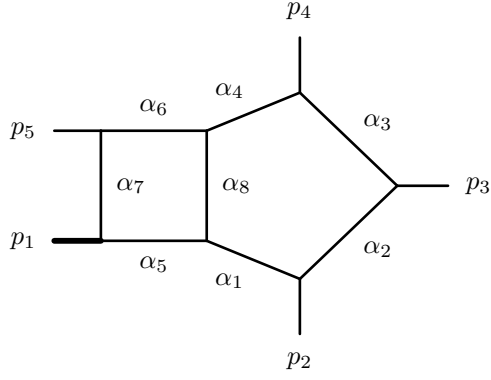
\begin{figure}
    \centering
    \input{Figures/5pt_1mass_zzz.tex}
\caption{A one-mass five-point pentabox family, with external momenta $p_i$ and internal propagators labeled by the corresponding Feynman parameters $\alpha_i$. Momentum $p_1$ is massive, as indicated by the bold line.}    \label{fig:5pt1masszzz}
\end{figure}

Let us illustrate how these algebraic branch points show up in a concrete example. Consider the one-mass five-point pentabox family shown in Figure~\ref{fig:5pt1masszzz}, which was computed in~\cite{Abreu:2020jxa}. One of the letters that appears in this integral is
\begin{align}
    W_{33}
    &=
    \frac{p_1^2-s_{23}+s_{45}+\sqrt{\Delta_3}}
         {p_1^2-s_{23}+s_{45}-\sqrt{\Delta_3}}\, ,
    \label{eq:w33-letter}
\end{align}
where
\begin{align}
    \Delta_3
    &=
    (p_1^2)^2+(s_{23}-s_{45})^2
    -2p_1^2(s_{23}+s_{45})\, .
    \label{eq:w33-delta3}
\end{align}
In this case,
\begin{equation}
    \bigl(p_1^2-s_{23}+s_{45}\bigr)^2-\Delta_3
    =
    4p_1^2s_{45}\, ,
    \label{eq:w33-factorization}
\end{equation}
so $W_{33}$ encodes three distinct singular loci---an algebraic branch point at $\Delta_3=0$, and logarithmic branch points at $p_1^2=0$ and $s_{45}=0$.

Using the propagator labeling shown in Figure~\ref{fig:5pt1masszzz}, the critical-point analysis associates the $\Delta_3=0$ singularity with four sectors:
\begin{equation}
\begin{aligned}
    S_{\Delta_3,1}&=\{1,3,7,8\}\, ,
    &\qquad
    S_{\Delta_3,2}&=\{3,5,7,8\}\, ,\\
    S_{\Delta_3,3}&=\{1,3,5,7,8\}\, ,
    &
    S_{\Delta_3,4}&=\{2,3,4,5,7,8\}\, .
\end{aligned}
\label{eq:delta3-sectors}
\end{equation}
Meanwhile, the $p_1^2=0$ singularity occurs in sector
$\{1,2,3,4,5,7,8\}$, while the $s_{45}=0$ singularity occurs in sector
$\{1,2,3,4,5,6,7,8\}$. Both of these sectors contain all of those in~\eqref{eq:delta3-sectors}, so the hierarchical principle allows a discontinuity with respect to either $p_1^2=0$ or $s_{45}=0$ to follow a discontinuity with respect to $\Delta_3=0$. 

By contrast, the symbol alphabet for this family of integrals also contains the letter $s_{23}+s_{34}$. As a singularity at $s_{23}+s_{34}=0$ only occurs in sector $\{2,3,5,6,8\}$, which contains none of the sectors in~\eqref{eq:delta3-sectors}, the hierarchical principle tells us that
\begin{equation}
    \text{Disc}_{s_{23}+s_{34}}\,
    \cdots\,
    \text{Disc}_{\Delta_3}\,
    \cdots\,\mathcal{I}
    =0\, .
\end{equation}
Consequently, any term containing the letter \(s_{23}+s_{34}\) must remain insensitive to the branch point at \(\Delta_3=0\) until the \(s_{23}+s_{34}=0\) singularity has been removed. In particular, this means that \(W_{33}\) (or any other letter that is inverted by the replacement \(\sqrt{\Delta_3}\to-\sqrt{\Delta_3}\)) cannot appear before \(s_{23}+s_{34}\), since computing a logarithmic discontinuity that removes $W_{33}$ would change the Galois signature of the symbol term (implying that the forbidden branch cut at \(\Delta_3=0\) was accessible either before or after this logarithmic discontinuity was computed).\footnote{If any of the algebraic letters in this family of integrals also involved a logarithmic branch point at $s_{23}+s_{34}=0$, this conclusion would be too quick, for the reasons given in the next paragraph; however, in this case, no such algebraic letters are present.}

Logarithmic branch points can also cancel between algebraic symbol letters. For instance, in the same pentabox family, two of the algebraic letters that appear are
\begin{equation}
    W_{44}
    =
    \frac{a_{44}+\sqrt{\Delta_5}}
         {a_{44}-\sqrt{\Delta_5}}\, ,
    \qquad
    W_{46}
    =
    \frac{a_{46}-\sqrt{\Delta_5}}
         {a_{46}+\sqrt{\Delta_5}}\, ,
\end{equation}
where
\begin{align}
    a_{44}
    &=
    -s_{12}s_{23}
    -s_{12}s_{51}
    +s_{23}s_{34}
    -s_{34}s_{45}
    +p_1^2s_{34}
    +s_{45}s_{51}\, , \\
    a_{46}
    &=
    s_{12}s_{23}
    -s_{12}s_{51}
    -s_{23}s_{34}
    +s_{34}s_{45}
    +2s_{34}s_{51}
    -p_1^2s_{34}
    +s_{45}s_{51}\, ,
\end{align}
and
\begin{equation}
    \Delta_5
    =
    \operatorname{Gram}(p_1,p_2,p_3,p_4)
    \label{eq:delta5-definition}
\end{equation}
is the five-point Gram determinant. It can be checked that both $W_{44}$ and $W_{46}$ give rise to a logarithmic branch point at
\begin{equation}
    s_{12}-s_{34}-s_{45}
    =0\, ,
\end{equation}
which implies they are subject to the following hierarchical constraint: 
\begin{align}
    \text{Disc}_{s_{12}-s_{34}-s_{45}}
    \cdots
    \text{Disc}_{p_1^2-s_{12}-s_{23}}
    \cdots \mathcal{I}
    =0\, .
    \label{pentabox_genalogical}
\end{align}
That is, a symbol term such as
\begin{equation}
    \cdots \otimes
    (p_1^2-s_{12}-s_{23})
    \otimes \cdots \otimes
    W_{44}
    \otimes \cdots
\end{equation}
would violate~\eqref{pentabox_genalogical}. However, on the support of $s_{12}-s_{34}-s_{45}=0$, $W_{46}$ and $W_{44}$ become identical, implying that the offending logarithmic branch point must cancel in the ratio $\frac{W_{44}}{W_{46}}$. Consequently, terms such as
\begin{equation} \label{eq:ratio_evading_hierarchical}
    \cdots \otimes
    (p_1^2-s_{12}-s_{23})
    \otimes \cdots \otimes
    \frac{W_{44}}{W_{46}}
    \otimes \cdots
\end{equation}
are allowed by~\eqref{pentabox_genalogical}.\footnote{Note that one cannot similarly avoid the constraint in~\eqref{pentabox_genalogical} by forming ratios of algebraic and rational letters, such as $W_{44}/(s_{12}{-}s_{34}{-}s_{45})$ or $W_{46}/(s_{12}{-}s_{34}{-}s_{45})$. For instance, consider a symbol term of the form 
\begin{equation*}
    \cdots \otimes
    (p_1^2-s_{12}-s_{23})
    \otimes \cdots \otimes
    \frac{W_{44}}{s_{12}{-}s_{34}{-}s_{45}}
    \otimes \cdots \, .
\end{equation*}
We can always isolate one of the letters that appear in the ratio by computing a discontinuity with respect to $\Delta_5=0$. For instance, if the part of the symbol represented by the ellipses is even with respect to the replacement $\sqrt{\Delta_5} \to - \sqrt{\Delta_5}$, we will get
\begin{align*}
    \text{Disc}_{\Delta_5} \Big(\cdots \otimes
    (p_1^2-s_{12}-s_{23})
    \otimes &\cdots \otimes
    \frac{W_{44}}{s_{12}{-}s_{34}{-}s_{45}}
    \otimes \cdots \Big) \\
    &= - 2 \big( \cdots \otimes
    (p_1^2-s_{12}-s_{23})
    \otimes \cdots \otimes
    W_{44}
    \otimes \cdots \big) \,.
\end{align*}
If, instead, the part of the symbol represented by the ellipses is odd with respect to this replacement, one will get 
\begin{align*}
    \text{Disc}_{\Delta_5} \Big(\cdots \otimes
    (p_1^2-s_{12}-s_{23})
    \otimes &\cdots \otimes
    \frac{W_{44}}{s_{12}{-}s_{34}{-}s_{45}}
    \otimes \cdots \Big) \\
    &=  2 \big( \cdots \otimes
    (p_1^2-s_{12}-s_{23})
    \otimes \cdots \otimes
    (s_{12}{-}s_{34}{-}s_{45})
    \otimes \cdots \big) \,.
\end{align*}
Either way, the remaining symbol term will violate~\eqref{pentabox_genalogical}. This argument also establishes why we can study the hierarchical constraints obeyed by rational symbol letters in isolation (as we did in Section~\ref{sec:examples})---algebraic letters cannot be used to hide the logarithmic branch points encoded by rational symbol letters in order to evade hierarchical constraints.} Thus, in the presence of algebraic letters, hierarchical constraints cannot always be imposed term by term in the symbol, since even logarithmic branch points can cancel between letters.

\begin{table}[t]
\centering
\small
\setlength{\tabcolsep}{5pt}
\renewcommand{\arraystretch}{1.3}
\begin{adjustbox}{max width=\textwidth}
\begin{tabular}{|c|c|c|c|c|}
\hline
\makecell{\textbf{Family of} \\ \textbf{Diagrams}}
& \makecell{\textbf{Mass} \\ \textbf{Configuration}}
& \makecell{\textbf{Genealogical} \\ \textbf{Constraints}}
& \makecell{\textbf{Symbol-Level} \\ \textbf{Restrictions}}
& \makecell{\textbf{Missed} \\ \textbf{ Constraints}} \\
\hline
\multirow{3}{*}{
\adjustbox{width=2.5cm,valign=c}{
\centering\begin{tikzpicture}[scale=1]
    \coordinate (a) at (-.8,0);
    \coordinate (b) at (0,.8);
    \coordinate (c) at (.8,0);
    \coordinate (d) at (0,-.8);
    \coordinate (e) at (2.2, 1.2);
    \coordinate (f) at (3, 0);
    \coordinate (g) at (2.2, -1.2);

    \coordinate (a1) at (-1.6, 0);
    \coordinate (c1) at (1.6, 0);
    \coordinate (e1) at (2.6, 1.6);
    \coordinate (f1) at (3.6, 0);
    \coordinate (g1) at (2.6, -1.6);

    \draw(a) -- (b) -- (c) -- (d) -- cycle;
    \draw (b) -- (e) -- (f) -- (g) -- (d);
    \draw[color=Maroon, line width = 3] (a1) -- (a);
    \draw(c1) -- (c);
    \draw(e1) -- (e);
    \draw (f1) -- (f);
    \draw (g1) -- (g);

\end{tikzpicture}
}}
& zzz & 62 & $547+12$ & 2 \\
\cline{2-5}
& mzz & 30 & $224+1$ & 0 \\
\cline{2-5}
& zmz & 88 & $575+11$ & 0 \\
\hline
\multirow{6}{*}{
\adjustbox{width=3.5cm,valign=c}{
\begin{tikzpicture}[scale=0.8]
    \coordinate (a) at (-.8,.8);
    \coordinate (b) at (.8,.8);
    \coordinate (c) at (.8,-.8);
    \coordinate (d) at (-.8,-.8);
    \coordinate (e) at (2.2, 1.2);
    \coordinate (f) at (3, 0);
    \coordinate (g) at (2.2, -1.2);

    \coordinate (a1) at (-1.6, 1.6);
    \coordinate (d1) at (-1.6, -1.6);
    \coordinate (e1) at (3, 2);
    \coordinate (f1) at (4, 0);
    \coordinate (g1) at (3, -2);

    \draw (a) -- (b) -- (c) -- (d) -- cycle;
    \draw (b) -- (e) -- (f) -- (g) -- (c);
    \draw[color=RoyalBlue, line width = 2] (a1) -- (a);
    \draw[color=Maroon, line width = 2] (d1) -- (d);
    \draw (e1) -- (e);
    \draw (f1) -- (f);
    \draw (g1) -- (g);
\end{tikzpicture}
}}
& zzz & 276 & $2266+36$ & 24 \\
\cline{2-5}
& zzm & 218 & $1693+69$ & 17 \\
\cline{2-5}
& zmz & 277 & $2257+58$ & 20 \\
\cline{2-5}
& mzz & 210 & $1629+35$ & 25 \\
\cline{2-5}
& mzm & 87 & $603+0$ & 0 \\
\cline{2-5}
& mmz & 159 & $1279+21$ & 31 \\
\hline
\end{tabular}
\end{adjustbox}
\caption{The additional genealogical constraints generated by \DiscKosky{} (beyond those reported in Table~\ref{tab:some 2 loop rational predictions}) once algebraic branch points are included, for the five-point one-mass hexabox and two-mass pentabox families. For each mass configuration, we report the number of genealogical constraints involving at least one algebraic branch point, and the number of restrictions they impose on the ordered letters (or linear combinations of letters) that can appear in the symbol. We decompose the latter number as $N_{\rm p}+N_{\rm ratio}$, where $N_{\rm p}$ counts the pairs of letters that are forbidden from ever appearing in a specific order, while $N_{\rm ratio}$ counts additional restrictions on which linear combinations of algebraic letters can appear before or after other letters, as in~\eqref{eq:ratio_evading_hierarchical}. These predictions have been checked to hold to all orders in dimensional regularization. However, unlike the constraints that \DiscKosky{} derives on the rational symbol letters in these integrals, this complete set of constraints does not seem to exhaust the full set of restrictions on which letters can follow each other in the symbol. In the last column, we report the number of additional restrictions of this type that can be observed in the differential equation matrices, beyond what follows from \DiscKosky{}'s predictions.}
\label{tab:mixed-algebraic-checks}
\end{table}

As these examples illustrate, the appearance of square roots in Feynman integrals gives rise to additional hierarchical constraints that further restrict how symbol letters can be combined in the final result. These algebraic constraints can be generated within \DiscKosky{} by supplying the relevant square-root arguments to \texttt{CheckSequentialDiscontinuity} (or the package's other functions) alongside an integral's logarithmic branch points. In practice, imposing these additional genealogical predictions requires considering possible cancellations between algebraic letters, or between algebraic letters and algebraic prefactors; consequently, these constraints must generally be imposed at the level of the full integral, rather than on individual symbol terms. Nevertheless, these constraints can be imposed mechanically, and we have checked the predictions made by \DiscKosky{} for the five-point one-mass hexabox and two-mass pentabox families of integrals. These predictions are summarized in Table~\ref{tab:mixed-algebraic-checks}. There, we see that an additional $1{,}407$ hierarchical constraints of the form~\eqref{eq:hierarchical-principle-background} can be derived once algebraic branch points are included. When combined with the constraints reported in Table~\ref{tab:some 2 loop rational predictions}, these exclude $14{,}195$ pairs of letters from appearing in a specific order in the symbol of these integrals (even if separated by other letters), and place an additional 243 restrictions on which linear combinations of algebraic letters can appear before or after other letters (again allowing any number of other letters in between) such that hierarchical constraints on the logarithmic branch points that appear in these letters can be evaded (as in the example involving $W_{44}$ and $W_{46}$ above). However, we cannot establish that these predictions exhaust the consequences of the hierarchical principle, as the differential equations reveal $119$ additional pairs of letters that never follow one another, beyond what is implied by \DiscKosky{}'s predictions. (Whether these additional restrictions actually follow from the hierarchical principle, or instead arise from some other combination of constraints, currently remains unclear.) Thus, we see that genealogical constraints---namely, those constraints that can be derived using the methods described in Section~\ref{sect:theoretical-background}---are not guaranteed to capture all restrictions of this type.

%% file: Figures/5pt_1mass_zzz.tex
\begin{tikzpicture}[
    scale=1.12,
    line cap=round,
    line join=round,
    ext/.style={line width=1pt},
    prop/.style={line width=1pt},
    massive/.style={line width=2.2pt}
]

    \coordinate (A) at (0,0.65);
    \coordinate (B) at (0,-0.65);
    \coordinate (C) at (1.25,0.65);
    \coordinate (D) at (1.25,-0.65);

    \coordinate (T) at (2.35,1.10);
    \coordinate (R) at (3.50,0);
    \coordinate (F) at (2.35,-1.10);

    \coordinate (P5) at (-0.55,0.65);
    \coordinate (P1) at (-0.55,-0.65);
    \coordinate (P4) at (2.35,1.75);
    \coordinate (P3) at (4.10,0);
    \coordinate (P2) at (2.35,-1.75);

    \draw[prop] (A) -- (C);
    \draw[prop] (B) -- (D);
    \draw[prop] (A) -- (B);
    \draw[prop] (C) -- (D);

    \draw[prop] (C) -- (T);
    \draw[prop] (T) -- (R);
    \draw[prop] (R) -- (F);
    \draw[prop] (F) -- (D);

    \draw[ext]     (P5) -- (A);
    \draw[massive] (P1) -- (B);
    \draw[ext]     (T) -- (P4);
    \draw[ext]     (R) -- (P3);
    \draw[ext]     (F) -- (P2);

    \node[left=3pt]  at (P5) {\small\textit{$p_5$}};
    \node[left=3pt]  at (P1) {\small\textit{$p_1$}};
    \node[above=3pt] at (P4) {\small\textit{$p_4$}};
    \node[right=3pt] at (P3) {\small\textit{$p_3$}};
    \node[below=3pt] at (P2) {\small\textit{$p_2$}};

    \node[above=2pt] at ($(A)!0.50!(C)$)
        {\small\textit{$\alpha_6$}};
    \node[below=2pt] at ($(B)!0.50!(D)$)
        {\small\textit{$\alpha_5$}};

    \node[right=2pt] at ($(A)!0.50!(B)$)
        {\small\textit{$\alpha_7$}};
    \node[right=2pt] at ($(C)!0.50!(D)$)
        {\small\textit{$\alpha_8$}};

    \node[above left=1pt] at ($(C)!0.53!(T)$)
        {\small\textit{$\alpha_4$}};
    \node[above right=1pt] at ($(T)!0.53!(R)$)
        {\small\textit{$\alpha_3$}};
    \node[below right=1pt] at ($(R)!0.47!(F)$)
        {\small\textit{$\alpha_2$}};
    \node[below left=1pt] at ($(F)!0.47!(D)$)
        {\small\textit{$\alpha_1$}};

\end{tikzpicture}

%% file: sections/Conclusions.tex
\section{Conclusions and Outlook}\label{sec:conclusions}

In this work, we introduced the \DiscKosky{} package, which automates the derivation of genealogical constraints on Feynman integrals (or, more generally, on Euler integrals). The package works by counting the solutions of the critical-point equations~\eqref{eq:critical-point-equations} sector by sector, both in generic kinematics and in candidate singular limits, to identify sectors in which the number of critical points decreases. Once the code has associated each singularity with a set of sectors, the hierarchical principle can be used to place restrictions on the sequential discontinuities that can appear in the final result. Internally, \DiscKosky{} computes the number of solutions to the critical point equations by constructing Gr\"obner bases over finite fields at randomly chosen kinematic points. This allows constraints to be derived in minutes, even for integrals at the current frontier of two- and three-loop analytic computations.

As part of our analysis, we have proposed a new test that allows us to identify discontinuities that cannot be computed more than once. Equipped with these additional predictions, we are able to capture the full consequences of the hierarchical principle in a number of examples, including the hook triangle, the two-mass easy box, and the one-mass five-point double box considered in Section~\ref{sec:non_repeating}. In each case, the underlying mechanism is the same---computing a discontinuity with respect to a singularity that arises in a sector containing no critical points (in generic kinematics) has the effect of putting additional propagators on shell, which places the original singularity out of reach.

The constraints derived by \DiscKosky{} turn out to be remarkably restrictive. Across the two-loop integral families considered in Tables~\ref{tab:some 2 loop rational predictions} and~\ref{tab:mixed-algebraic-checks}, \DiscKosky{} generates a total of $4{,}529$ genealogical constraints. Jointly, these discontinuity restrictions forbid $14{,}195$ pairs of symbol letters from appearing in a given order (even when separated by other letters), out of $51{,}711$ possible pairs. An additional 243 restrictions can also be placed on which linear combinations of algebraic letters can appear before or after other letters. All of these predictions have been verified against the known canonical differential equations, and hold to all orders in dimensional regularization. Moreover, these predictions account for nearly all restrictions of this type (as observed in the differential equations), missing only $119$ cases in which specific pairs of symbol letters never appear in a given order.

In this paper, we have focused on examples that evaluate to multiple polylogarithms. Starting already at two loops, however, Feynman integrals can evaluate to more complicated functions that involve (for instance) curves of high genus or Calabi--Yau manifolds~\cite{Laporta:2004rb,Adams:2013kgc,Bloch:2013tra,Huang:2013kh,Adams:2014vja,Bloch:2014qca,Remiddi:2016gno,Broedel:2017kkb,Marzucca:2023gto,Bourjaily:2018ycu,Bourjaily:2018yfy,Klemm:2019dbm,Bourjaily:2019hmc,Bonisch:2021yfw,Jockers:2024uan,Bargiela:2025vwl}. \DiscKosky{} can also be used to derive genealogical constraints that restrict the discontinuity structure of these more complicated integrals. More generally, when used in conjunction with the tools that now exist for locating the singularities of Feynman integrals~\cite{Fevola:2023fzn,Fevola:2023kaw,Correia_2026,Chestnov:2026mpo}, \DiscKosky{} forms part of an increasingly reliable pipeline that can be fed into bootstrap approaches for computing Feynman integrals, or into the construction of differential equations. (We emphasize that whenever the algorithm presented in Section~\ref{sect:theoretical-background} cannot be applied, \DiscKosky{} returns an \texttt{Indeterminate} result, ensuring that no false predictions are made.) Understanding more concretely how genealogical constraints can be exploited in bootstrap approaches to elliptic and Calabi--Yau integrals constitutes an interesting direction for future work. 

The methods implemented in \DiscKosky{} can also be used beyond the context of Feynman integrals. This was illustrated in Section~\ref{sec:correlators}, where we applied the code to integrals involving linear denominators and generic twist exponents. While the twist exponents obscure which sector each critical point originates in, this sector information can nevertheless be reconstructed and used to derive genealogical constraints~\cite{Frellesvig:2019uqt,Frellesvig:2020qot,Brunello:2023rpq}. Using these ideas, \DiscKosky{} can be applied to broad classes of Euler integrals.

There remain, however, important classes of integrals for which the present method is not yet sufficient. In particular, more work is needed to extend this method to cases in which higher-dimensional solutions to the critical-point equations exist in generic kinematics, or in which the solutions to these equations fail to account for all admissible integration cycles~\cite{Boels:2015yna,Bitoun:2017nre,Crisanti:2026rbc}. One possible approach would be to resolve these critical varieties into isolated contributions; another would be to formulate the analysis more directly in terms of the relative homology of admissible cycles. Either approach would extend genealogical predictions to a broader class of integrals, and could lead to a more systematic understanding of magic relations~\cite{Crisanti:2026rbc}.

By virtue of centering our analysis around the hierarchical principle, this paper has focused on identifying pairs of discontinuities that cannot follow one another in Feynman integrals, even if additional discontinuities are computed in between. However, further constraints on the discontinuities of Feynman integrals are also known to exist, such as the Steinmann relations, which restrict which discontinuities can be computed in immediate succession. Given the extraordinary constraining power that the (extended) Steinmann relations have in planar \(\mathcal{N}=4\) super-Yang--Mills theory~\cite{Caron-Huot:2016owq,Dixon:2016nkn,Drummond:2017ssj,Caron-Huot:2019bsq}, it would be highly desirable to develop a method for predicting when such adjacency restrictions hold between the discontinuities of individual Feynman integrals.

Finally, the constraints derived here apply integral by integral, and as such cannot capture analytic properties that only emerge at the level of amplitudes through nontrivial cancellations between integrals. Understanding how to predict these additional, theory-dependent properties represents an important step towards applying these methods to phenomenologically relevant processes. Progress in this direction may require new representations of amplitudes that more directly align their analytic structure with that of the underlying integrals, making cancellations and inherited constraints easier to track. Finding representations in which these cancellations are manifest could provide a practical route toward extending the genealogical methods developed here to full amplitudes.